\documentclass{ws-ijgmmp}

\usepackage{cite}
\usepackage[colorlinks=true,linkcolor=red,
            urlcolor=gray,
            citecolor=blue]{hyperref}

\newcommand{\dd}{\mathrm{d}}
\newcommand{\Ri}{\mathcal{R}}
\usepackage{color}
\usepackage{xcolor}
\usepackage{soul}
\usepackage{tensor}
\def\tns{\tensor}

\definecolor{tp}{RGB}{148,0,211}

\newcommand{\beq}{\begin{equation}}
\newcommand{\eeq}{\end{equation}}

\newcommand{\bet}{\beta}

\newcommand{\eps}{\epsilon}

\newcommand{\lam}{\lambda}
\newcommand{\vphi}{\varphi}
\DeclareMathOperator{\arcsinh}{arcsinh}

\begin{document}

\markboth{I.~D.~GIALAMAS, A.~KARAM, T.~D.~PAPPAS \& E.~TOMBERG}
{Implications of Palatini gravity for inflation and beyond}

\catchline{}{}{}{}{}

\title{IMPLICATIONS OF PALATINI GRAVITY \\
FOR INFLATION AND BEYOND
}

\author{IOANNIS~D.~GIALAMAS}

\address{Laboratory of High Energy and Computational Physics, \\
National Institute of Chemical Physics and Biophysics, \\R{\"a}vala pst.~10,  Tallinn, 10143, Estonia\\
\email{ \textit{ioannis.gialamas@kbfi.ee} } }

\author{ALEXANDROS KARAM}

\address{Laboratory of High Energy and Computational Physics, \\
National Institute of Chemical Physics and Biophysics, \\R{\"a}vala pst.~10,  Tallinn, 10143, Estonia\\
\email{ \textit{alexandros.karam@kbfi.ee} } }

\author{THOMAS~D.~PAPPAS}
\address{Research Centre for Theoretical Physics and Astrophysics,\\
Institute of Physics, Silesian University in Opava,\\
Bezručovo nám.~13, CZ-746 01 Opava, Czech Republic\\
\email{\textit{thomas.pappas@physics.slu.cz}}}

\author{EEMELI TOMBERG}

\address{Laboratory of High Energy and Computational Physics, \\
National Institute of Chemical Physics and Biophysics, \\R{\"a}vala pst.~10,  Tallinn, 10143, Estonia\\
\email{ \textit{eemeli.tomberg@kbfi.ee} } }

\maketitle

\begin{abstract}

We present an introduction to cosmic inflation in the framework of Palatini gravity, which provides an intriguing alternative to the conventional metric formulation of gravity. In the latter, only the metric specifies the spacetime geometry, whereas in the former, the metric and the spacetime connection are independent variables---an option that can result in a gravity theory distinct from the metric one. In scenarios where the field(s) responsible for cosmic inflation are non-minimally coupled to gravity or the gravitational sector is extended, assumptions about the underlying gravitational degrees of freedom can have substantial implications for the observational effects of inflation. We examine this explicitly by discussing various compelling scenarios, such as Higgs inflation with a non-minimal coupling to gravity, Higgs inflation with a non-minimal derivative coupling, $\Ri^2$ inflation, and beyond. We also comment on reheating in these models. Finally, as an application of the general results of Palatini $\Ri^2$ inflation, we review a model of successful quintessential inflation, where a single scalar field acts initially as the inflaton and then becomes dynamical dark energy, in agreement will all experimental constraints.

\end{abstract}

\keywords{Cosmic inflation; Palatini gravity; Review.}

\begingroup
 \hypersetup{linkcolor=black} 
 \tableofcontents 
\endgroup

\section{Introduction}

The concordance cosmological model, also known as $\Lambda$ Cold Dark Matter ($\Lambda$CDM), provides a comprehensive explanation for the content and evolution of the Universe. The model consists of three major components---ordinary matter (which includes photons and neutrinos), cold dark matter (CDM), and the cosmological constant $\Lambda$.
The $\Lambda$CDM model also incorporates two phases of accelerated expansion that occurred in the early and late Universe, known as cosmic inflation and dark energy. Overall, this model provides a robust framework for understanding the Universe and its history.

The inflationary epoch~\cite{Starobinsky:1980te, Kazanas:1980tx, Sato:1980yn, Guth:1980zm, Linde:1981mu, Albrecht:1982wi, Linde:1983gd} was initially proposed to address the limitations of the hot Big Bang model. Inflation can create a flat, uniform, and isotropic Universe without topological defects, matching observations. Additionally, it offers a natural explanation for how quantum fluctuations of gravitational and matter fields in the vacuum can be amplified to cosmological perturbations~\cite{Starobinsky:1979ty, Mukhanov:1981xt, Hawking:1982cz, Starobinsky:1982ee, Guth:1982ec, Bardeen:1983qw}. These perturbations later serve as the seeds for the primordial anisotropy in the cosmic microwave background (CMB) and the formation of the Universe's large-scale structure. The most basic form of inflation involves a scalar field that is minimally coupled to gravity, featuring a canonical kinetic term and controlled by a potential energy density that initiated the exponential expansion of the Universe. Nonetheless, recent findings by the Planck~\cite{Planck:2018jri} and BICEP/Keck~\cite{BICEP:2021xfz} collaborations have invalidated many simple models of inflation, such as monomial chaotic inflation or natural inflation, which has led to the examination of more intricate models. According to the constraint on the tensor-to-scalar ratio ($r<0.036$)~\cite{BICEP:2021xfz}, the inflaton potential must be very flat at large field values. To accomplish this, one straightforward approach is to couple the inflaton field non-minimally to gravity or introduce a quadratic curvature term in the action. These terms arise from quantum corrections, making their inclusion natural.

In the presence of a non-minimal gravity sector, the exact formulation of general relativity becomes important. A form of modified gravity, the Palatini (or metric-affine) formulation~\cite{Palatini1919, Ferraris1982} has become increasingly popular as an alternative to the standard metric formulation\footnote{For more details about modified gravity in both formulations we point the reader to~\cite{Sotiriou:2006qn, Olmo:2011uz} for reviews in the Palatini formulation and to~\cite{Sotiriou:2006hs, Sotiriou:2008rp, Nojiri:2010wj, DeFelice:2010aj, Clifton:2011jh, Capozziello:2011et, Nojiri:2017ncd} for reviews in the metric formulation. Also, the cosmological perturbations in the Palatini formulation have been studied in~\cite{Koivisto:2005yc} and recently in~\cite{Kubota:2020ehu}.}. In the Palatini approach, the metric tensor $g_{\mu\nu}$ and the connection $\tns{\Gamma}{^\rho_\mu_\nu}$ are treated as independent variables, and the action is varied with respect to both. When the scalar field is minimally coupled and the action is linear in the Ricci scalar $\Ri$, the Palatini and metric formulations yield the same equations of motion, and the connection assumes the Levi-Civita form. However, if the field is non-minimally coupled to gravity~\cite{Bauer:2008zj, Bauer:2010bu, Tamanini:2010uq, Bauer:2010jg, Rasanen:2017ivk, Tenkanen:2017jih, Racioppi:2017spw, Markkanen:2017tun, Jarv:2017azx, Fu:2017iqg, Racioppi:2018zoy, Carrilho:2018ffi, Kozak:2018vlp, Rasanen:2018fom, Rasanen:2018ihz, Almeida:2018oid, Shimada:2018lnm, Takahashi:2018brt, Jinno:2018jei, Rubio:2019ypq, Tenkanen:2019xzn, Bostan:2019uvv, Bostan:2019wsd,  Racioppi:2019jsp, Tenkanen:2020dge, Shaposhnikov:2020fdv, Borowiec:2020lfx, Jarv:2020qqm, Karam:2020rpa, McDonald:2020lpz, Langvik:2020nrs, Shaposhnikov:2020gts, Shaposhnikov:2020frq, Mikura:2020qhc, Gialamas:2020vto, Verner:2020gfa, Enckell:2020lvn, Reyimuaji:2020goi, Karam:2021wzz,  Mikura:2021ldx, Gomez:2021roj, Racioppi:2021ynx,  Mikura:2021clt, Cheong:2021kyc, Azri:2021uat, Racioppi:2021jai, Karananas:2022byw, Rigouzzo:2022yan, Panda:2022esd, Yin:2022fgo, Gialamas:2022gxv} and/or quadratic or higher curvature terms are added~\cite{Meng:2004yf, Borunda:2008kf, Bombacigno:2018tyw, Enckell:2018hmo, Antoniadis:2018ywb, Antoniadis:2018yfq, Tenkanen:2019jiq, Edery:2019txq, Giovannini:2019mgk, Tenkanen:2019wsd, Gialamas:2019nly, Antoniadis:2019jnz, Tenkanen:2020cvw, Lloyd-Stubbs:2020pvx, Antoniadis:2020dfq,  Ghilencea:2020piz, Das:2020kff, Gialamas:2020snr, Ghilencea:2020rxc, Iosifidis:2020dck, Bekov:2020dww, Gomez:2020rnq, Dimopoulos:2020pas,  Karam:2021sno, Lykkas:2021vax, Gialamas:2021enw, Antoniadis:2021axu, Annala:2021zdt, Gialamas:2021rpr, Giovannini:2021due, AlHallak:2021hwb,  Dioguardi:2021fmr, Dimopoulos:2022tvn, Pradisi:2022nmh, Durrer:2022emo, Salvio:2022suk,  Lahanas:2022mng, Gialamas:2022xtt, Gialamas:2023aim}, the Palatini connection deviates from Levi-Civita, and inflationary models in the two formulations give different predictions \textit{e.g.} for the CMB observables. In practice, the first step in inflationary computations is to Weyl transform the metric tensor so that the theory is reduced back to a minimal form, going from the original Jordan frame to the new Einstein frame. Since the Riemann tensor depends on the connection, $\Ri$ transforms differently in the two formulations, leading to different results.

More formally, the Levi-Civita connection $\tns{\widetilde{\Gamma}}{^\rho_\mu_\nu}$ is a function of the metric, given by
\begin{equation}
\label{eq:Levi_Civ}
\tns{\widetilde{\Gamma}}{^\rho_\mu_\nu} \equiv \frac{1}{2}g^{\rho\sigma}\left(\partial_\mu g_{\sigma\nu} +\partial_\nu g_{\sigma\mu} -\partial_\sigma g_{\mu\nu} \right)\,.
\end{equation}
The general Palatini connection $\tns{\Gamma}{^\rho_\mu_\nu}$ can be decomposed as
\begin{equation}
\label{eq:gen_Con}
\tns{\Gamma}{^\rho_\mu_\nu} = \tns{\widetilde{\Gamma}}{^\rho_\mu_\nu} + \tns{C}{^\rho_\mu_\nu}\,,
\end{equation}
where $\tns{C}{^\rho_\mu_\nu}$ is the distortion tensor. The Riemann tensor is defined as
\begin{equation}
\label{eq:Riemann}
\tns{\mathcal{R}}{^\rho_\sigma_\mu_\nu} \equiv \partial_\mu \tns{\Gamma}{^\rho_\nu_\sigma} - \partial_\nu \tns{\Gamma}{^\rho_\mu_\sigma} + \tns{\Gamma}{^\rho_\mu_\lambda} \tns{\Gamma}{^\lambda_\nu_\sigma} - \tns{\Gamma}{^\rho_\nu_\lambda} \tns{\Gamma}{^\lambda_\mu_\sigma}\,,
\end{equation}
for which (unlike the usual metric Riemann tensor) the only symmetry is the antisymmetry under the interchange of the last two indices. As a result of the complexity of $\tns{\mathcal{R}}{^\rho_\sigma_\mu_\nu}$, it has three non-zero contractions,
\begin{equation}
\label{eq:Ricci_tensor}
    \mathcal{R}_{\mu\nu} \equiv \tns{\mathcal{R}}{^\rho_\mu_\rho_\nu}\,, \quad \tns{\widehat{\mathcal{R}}}{^\mu_\nu}\equiv g^{\rho\sigma}\tns{\mathcal{R}}{^\mu_\rho_\sigma_\nu}\,, \quad \mathcal{R}'_{\mu\nu} \equiv \tns{\mathcal{R}}{^\rho_\rho_\mu_\nu}\,,
\end{equation}
called the Ricci, co-Ricci, and homothetic curvature tensor, respectively. The Ricci scalar is uniquely defined from a further contraction of the Ricci or the co-Ricci tensor as
\begin{equation}
\label{eq:Ricci_scalar}
    \mathcal{R} \equiv g^{\mu\nu} \mathcal{R}_{\mu\nu} = - \tns{\widehat{\mathcal{R}}}{^\mu_\mu}\,.
\end{equation}
As the homothetic curvature is an antisymmetric 2-rank tensor, its contraction with the symmetric metric tensor vanishes, so no additional scalar arises\footnote{In Palatini gravity there is also an additional scalar which is linear in the Riemann tensor. It is called the Holst invariant~\cite{Holst:1995pc} and is defined as $\widetilde{\mathcal{R}} = \sqrt{-g} \epsilon^{\mu\nu\rho\sigma}\mathcal{R}_{\mu\nu\rho\sigma}$, with $g$ being the determinant of the metric tensor and $\epsilon^{\mu\nu\rho\sigma}$ the totally antisymmetric Levi-Civita symbol, with $\epsilon^{0123}=1$. In the metric formulation it vanishes identically due to the symmetry properties of the Riemann tensor. Recently, interest has been rekindled on the study of the Holst invariant and its influence on inflationary dynamics~\cite{Langvik:2020nrs, Shaposhnikov:2020gts, Shaposhnikov:2020frq, Pradisi:2022nmh, Salvio:2022suk, Gialamas:2022xtt}. }.

Below, we will compare $\mathcal{R}$ to $R$, defined as the Ricci scalar computed from the Levi-Civita connection \eqref{eq:Levi_Civ} and depending only on the metric. These two are related by
\begin{equation}
\label{eq:ricci_C}
\mathcal{R}=R+\widetilde{\nabla}_{ \mu}\tns{C}{_\nu^\mu^\nu}-\widetilde{\nabla}_{ \nu}\tns{C}{_\mu^\mu^\nu}+\tns{C}{_\mu^\mu_\lambda}\tns{C}{_\nu^\lambda^\nu}-\tns{C}{_\nu^\mu_\lambda}\tns{C}{_\mu^\lambda^\nu}\,,
\end{equation}
where $\widetilde{\nabla}_\mu$ is the covariant derivative taken with $\tns{\widetilde{\Gamma}}{^\rho_\mu_\nu}$. In the metric formulation of gravity, $\tns{C}{^\rho_\mu_\nu}=0$, $\tns{\widetilde{\Gamma}}{^\rho_\mu_\nu}=\tns{\Gamma}{^\rho_\mu_\nu}$, and $\Ri = R$.

In many ways, the Palatini formalism is better-behaved and more natural than its metric counterpart. For one, it contains fewer assumptions about the relationships between the gravitational degrees of freedom. In addition, the Riemann tensor \eqref{eq:Riemann}---and thus typical gravity actions---only contain first derivatives of $\tns{\Gamma}{^\rho_\mu_\nu}$ in the Palatini formulation, contrary to its metric counterpart with second derivatives of $g_{\mu\nu}$ through $\partial_\mu \tns{\widetilde{\Gamma}}{^\rho_\nu_\sigma}$ terms. Due to this, careful computations in the metric formulation require adding the York--Gibbons--Hawking boundary term \cite{York:1972sj, Gibbons:1976ue} to produce the correct equations of motion. This term is unnecessary in the Palatini case. In addition, the Palatini formulation of Higgs inflation preserves tree-level unitarity better than its metric counterpart \cite{Barbon2009,Bauer:2010jg,Shaposhnikov:2020geh, McDonald:2020lpz, Mikura:2021clt, Ito:2021ssc, Antoniadis:2021axu, Panda:2022esd}, making the model better behaved from a quantum field theoretic point of view. These features make the Palatini formulation theoretically appealing.

In this review, we will showcase models of Palatini inflation with non-minimal gravity sectors. They demonstrate the rich model-building potential of Palatini inflation and highlight the CMB predictions of this scenario, such as the tendency for very small values of the tensor-to-scalar ratio $r$. Such models make inflation a viable early-Universe framework even if tensor perturbations are not detected by next-generation CMB experiments. On the other hand, by distinguishing the predictions of the Palatini models from their metric counterparts, CMB observations can, in principle, probe the fundamental degrees of freedom for gravity. This paper does not aim to discuss every Palatini model of inflation found in the literature in detail; instead, we introduce some central concepts and results through pedagogical examples and point out further references to the interested reader.

The review is structured as follows. In section~\ref{sec:minimal_inflation}, we discuss single-field slow-roll inflation and its CMB predictions in the standard metric setup and introduce the concept of a Weyl transformation through examples. In section~\ref{sec:scalar_tensor}, we compare the metric and Palatini formulations in scalar-tensor theories, where the gravity action is linear in $\Ri$, but it may couple non-minimally to the inflaton scalar. In sections~\ref{sec:non_min_derivative_coupling} and~\ref{sec:quadratic_gravity}, we expand the discussion by including a non-minimal derivative coupling and second-order curvature terms, respectively. Sections~\ref{sec:preheating} and~\ref{sec:quintessece} go beyond inflation, considering reheating---particle production heralding the transition from inflation to the radiation-dominated epoch---and combining inflation with quintessence in a model where the same scalar field causes the late-Universe acceleration.

Throughout the review, we adopt the $(-,+,+,+)$ signature for the metric.
We denote the canonical Jordan-frame (JF) and Einstein-frame (EF) fields with $\vphi$ and $\phi$, respectively, and reserve $\chi$ for an auxiliary field used at intermediary steps. Finally, we use an overbar to denote the EF metric $\bar{g}_{\mu\nu}$ and its associated quantities such as $\bar{\nabla}_{\mu}$. We use natural units, setting the reduced Planck mass $M_{\rm Pl}$ to one.

\section{Minimal inflation}
\label{sec:minimal_inflation}

Before diving into the world of Palatini gravity, let us review inflationary model building in a standard setup. A canonical model of metric single-field inflation is defined by the standard action 
\begin{equation} \label{eq:canonical_S}
S = \int \dd^4 x \sqrt{- g} \left[ \frac{R}{2} - \frac{1}{2} g^{\mu\nu} \partial_\mu \phi \partial_\nu \phi - U(\phi) \right] \,,
\end{equation}
where $g$ is the metric determinant, $R$ is the Ricci scalar, $\phi$ is the inflaton field, and $U(\phi)$ is its potential.

During inflation, the Universe is homogeneous and isotropic, with the FRW metric $\dd s^2 = -\dd t^2 + a^2(t)\dd x_i^2$, where $t$ is the cosmic time, $x_i$ are the spatial coordinates, and $a(t)$ is the scale factor. All dynamic quantities only depend on $t$. Varying \eqref{eq:canonical_S} with respect to the metric components yields the Friedmann equations:
\begin{equation} \label{eq:Friedmann}
\left( \frac{\dot{a}}{a} \right)^2 \equiv H^2 = \frac{1}{3} \left[ \frac{\dot{\phi}^2}{2} + U \right] \, , \qquad \dot{H} = - \frac{1}{2} \dot{\phi}^2 \, ,
\end{equation}
where a dot denotes a derivative with respect to $t$, and $H$ is the Hubble parameter. Varying \eqref{eq:canonical_S} with respect to $\phi$ yields the useful, though not independent, Klein--Gordon equation:
\begin{equation} \label{eq:klein_gordon}
\ddot{\phi} + 3 H \dot{\phi} + U' = 0 \, ,
\end{equation}
where a prime denotes differentiation with respect to the inflaton field $\phi$. During inflation, $\phi$ rolls down the slope of the potential $U$, slowed down by the friction term $3 H \dot{\phi}$.

\subsection{Slow-roll approximation and CMB observables}
\label{sec:SR_and_CMB}
To characterize the behaviour of $\phi$, we define the Hubble slow-roll parameters
\begin{equation} \label{eq:sr_parameters}
\epsilon_H \equiv - \frac{\dot{H}}{H^2} = \frac{3 \dot{\phi}^2}{\dot{\phi}^2 + 2 U} \,, \quad \eta_H \equiv - \frac{\ddot{\phi}}{H \dot{\phi}} \, .
\end{equation}
The first parameter obeys $\frac{\ddot{a}}{a} = H^2 (1 - \epsilon_H)$, so $\eps_H < 1$ is a sufficient and necessary condition for inflation. In other words, inflation takes place when the potential dominates over the kinetic energy. The second parameter $\eta_H$ characterizes the sizes of the different terms in \eqref{eq:klein_gordon}. In the slow-roll (SR) limit, both parameters are small, so that
\begin{equation} \label{eq:sr_limit}
U(\phi) \gg \dot{\phi}^2  \,, \qquad  \vert \ddot{\phi} \vert \ll \vert 3 H \dot{\phi} \vert , \, \vert U' \vert \, ,
\end{equation}
and the equations of motion simplify to
\begin{equation} \label{eq:sr_eoms}
H^2 \approx  \frac{1}{3} U(\phi) \,, \quad
\dot{\phi} \approx  - \frac{U'}{3 H} \, .
\end{equation}
The shape of the potential is encoded in the potential slow-roll parameters
\begin{equation} 
\epsilon_U \equiv \frac{1}{2} \left( \frac{U'}{U} \right)^2 \, , \qquad  \eta_U \equiv \frac{U''}{U} \, .
\end{equation}
If both of these are small, slow-roll is an attractor: Hubble friction decreases the kinetic energy of the field until the friction and the potential push are balanced and \eqref{eq:sr_limit} applies. In the slow-roll limit, $\epsilon_H\approx\epsilon_U$, $\eta_H \approx \eta_U - \epsilon_U$.

\begin{figure}
\centering
\includegraphics[width=0.75\textwidth]{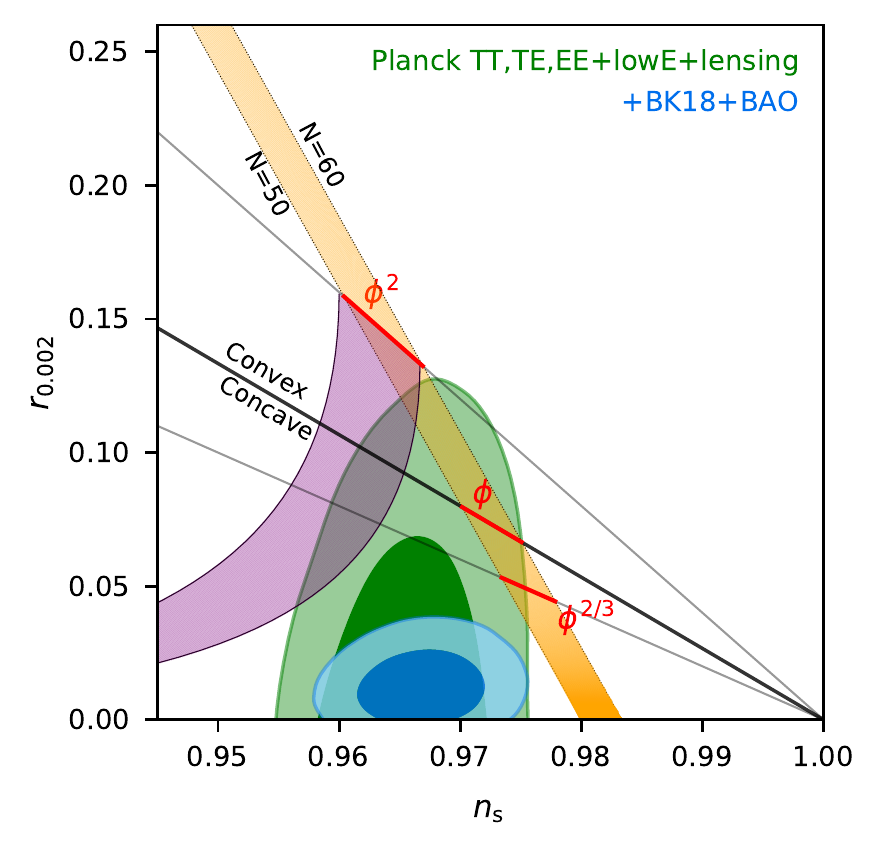}
\caption{Latest observational constraints on the CMB observables $n_s$ and $r$ from a combination of probes. The light and dark regions correspond to $2\sigma$ and $1\sigma$ confidence. Predictions of three chaotic potentials have been plotted (see section~\ref{sec:chaotic_inflation}) together with a line separating concave inflationary potentials from convex ones. The yellow-shaded region corresponds to $N_\star = 50 - 60$ $e$-folds for the chaotic models, while the purple-shaded region corresponds to the predictions of natural inflation. Figure adapted from~\cite{BICEP:2021xfz}.}
\label{fig:nsr_cons}
\end{figure}

Cosmological perturbations are described by the curvature perturbation $\zeta$, imprinted on the CMB radiation. CMB observations show that, around the pivot scale $k_\star=0.05\, {\rm Mpc}^{-1}$, the perturbations are very close to Gaussian, with a power-law power spectrum~\cite{Planck:2018jri}
\begin{equation}
\label{eq:power_spectrum_observations}
\mathcal{P}_{\zeta}(k) \equiv A_s \left(\frac{k}{k_\star}\right)^{n_s-1}\,, \quad
A_s \simeq 2.1\times 10^{-9} \,, \quad
n_s = 0.9649 \pm 0.0042 \, ,
\end{equation}
where $n_s-1= \dd \ln \mathcal{P}_{\zeta}(k) /\dd \ln k$.
The perturbations have their origin in inflation. Slow-roll inflation, in particular, produces a nearly scale-invariant spectrum ($n_s \approx 1$) compatible with observations. The slow-roll predictions for the spectral strength and tilt are (see \textit{e.g.} \cite{Lyth:2009zz})
\begin{equation} \label{eq:power_spectrum_sr}
A_s \equiv \frac{1}{24 \pi^2 } \frac{U(\phi_\star)}{\epsilon_U(\phi_\star)} \, , \qquad
n_s -1 \simeq -6\epsilon_U + 2\eta_U\,,
\end{equation}
where $\phi_\star$ is the field value when the pivot scale left the Hubble radius, $k_\star=aH$. Similarly, the power spectrum of primordial gravitational waves and the tensor-to-scalar ratio read~\cite{Planck:2018jri,BICEP:2021xfz}
\begin{equation} \label{eq:r}
\mathcal{P}_{T} \equiv  8 \left( \frac{H}{2\pi}\right)^2 \simeq \frac{2 U}{3 \pi^2} \, , \qquad
r \equiv \frac{\mathcal{P}_{T}}{\mathcal{P}_{\zeta}} \simeq 16\epsilon_U < 0.036 \, ,
\end{equation}
where the last limit again comes from CMB observations~\cite{Planck:2018jri,BICEP:2021xfz}.  The tensor-to-scalar ratio is usually calculated at the pivot scale $k_\star = 0.002\, \mbox{Mpc}^{-1}$ in order to compare with the Planck contours shown in figure~\ref{fig:nsr_cons}, while the bound of equation~\eqref{eq:r} applies at $k_\star = 0.05\, \mbox{Mpc}^{-1}$. 

To find the pivot scale field value $\phi_\star$, we relate it to the number of $e$-foldss $N_\star$ (defined as $N=\ln a$) between the Hubble exit of $k_\star$ and the end of inflation. To first order in slow roll,
\begin{equation} \label{eq:N_SR}
    N_\star \simeq \int_{\phi_{\rm end}}^{\phi_\star} \frac{{\rm d} \phi}{\sqrt{2\epsilon_U}} \, ,
\end{equation}
where $\phi_{\rm end}$ is the field value at the end of inflation, near the potential minimum, determined from $\epsilon_H \simeq \epsilon_U = 1$. Typically,  $N_\star=50$--$60$, depending on the energy scale of inflation and the post-inflationary expansion history. Below, when we need a more accurate result, we use \cite{Liddle:2003as}
\begin{eqnarray}
    N_\star =& \ln \left[ \left(\frac{\pi^2}{30} \right)^{\frac{1}{4}} \frac{(g_{\star\, s}(T_0))^{ \frac{1}{3}}}{ \sqrt{3} } \frac{T_0}{H_0} \right]
 - \ln\left[ \frac{k_\star}{  a_0 H_0} \right] 
 + \frac{1}{4} \ln\left[\frac{U^2(\phi_\star)}{\rho_{\rm end}} \right] \nonumber\\
+&  \dfrac{1- 3 w}{12 ( 1+w)}  \ln  \frac{\rho_{\rm reh} }{ \rho_{\rm end }} 
 + \frac{1}{4} \ln \left[ \frac{ g_{\star}(T_{\rm reh})}{ g_{\star\, s}(T_{\rm reh})  } \right]  - \frac{1 }{12}  \ln  \left[ g_{\star\, s}(T_{\rm reh}) \right]\,.
 \label{eq:efolds}
\end{eqnarray}
Here the subscripts denote quantities calculated at the present epoch ($ ``0" $), at reheating ($``{\rm reh}"$), and at the end of inflation ($``{\rm end}"$). $T$ is the temperature of the Universe, with $T_0 \approx 2.7\,\mathrm{K}$. The entropy density degrees of freedom are denoted by $g_{\star\, s}$, with $g_{\star\, s}(T_0)=43/11$ today and $g_{\star\, s}(T_{\rm reh}) \simeq  \mathcal{O}(10^2)$ during the reheating phase for $T_{\rm reh}\sim 1\,\mathrm{TeV}$ or higher.
The energy density degrees of freedom are also  $g_{\star}(T_{\rm reh}) =  \mathcal{O}(10^2)$. The inflaton's energy density is given by $\rho = \frac{1}{2}\dot{\phi}^2 + U(\phi)$. Furthermore, $w$ is the equation-of-state parameter during reheating which in this review is taken to be $w=1/3$, \textit{i.e.} the Universe's thermalization  is instantaneous. Using $U(\phi_\star) \simeq \frac{3\pi^2}{2}A_s r$ and $\rho_{\rm end} = \frac{3}{2} U(\phi_{\rm end})$, equation~\eqref{eq:efolds} simplifies to
\begin{equation}
N_{\star} \simeq 55.8 + \frac{1}{4} \ln \left[\frac{r}{0.036} \right] -\frac{1}{4} \ln \left[\frac{U(\phi_{\rm end})}{U(\phi_{\star})} \right]\quad \text{for} \quad k_\star = 0.05\, \mbox{Mpc}^{-1}\,.
\label{eq:e-folds:simplified}
\end{equation}

The observations and predictions \eqref{eq:power_spectrum_observations}--\eqref{eq:N_SR} guide inflationary model building, linking the CMB to the inflaton potential. This is our starting point for analyzing the different inflationary models below. Figure~\ref{fig:nsr_cons} plots the predictions of example models against the CMB constraints~\cite{Planck:2018jri,BICEP:2021xfz}.

\subsection{Chaotic inflation}
\label{sec:chaotic_inflation}
As a demonstration of the above machinery, we consider chaotic inflation~\cite{Linde:1983gd}, defined by the potential
\begin{equation}
U(\phi) = \lambda_n \phi^n \, .
\end{equation}
The first two potential slow-roll parameters are now
\begin{equation}
\epsilon_U = \frac{1}{2} \left( \frac{U'}{U} \right)^2 = \frac{n^2}{2} \frac{1}{\phi^2} \, , \qquad
\eta_U = \frac{U''}{U} = n \left( n - 1 \right) \frac{1}{\phi^2} \, .
\end{equation}
We then have 
\begin{equation} 
\phi_{\mathrm{end}} = n/\sqrt{2} \quad \Rightarrow \quad N_\star = \frac{\phi^2_\star}{2 n} - \frac{n}{4} \quad \Rightarrow \quad \phi_\star^2 = 2 n N_\star + \frac{n^2}{2}
\end{equation}
and, writing the SR parameters in terms of $N_\star$,
\begin{equation} 
n_s = 1 - \frac{2 n + 4 }{4 N_\star + 1} \, , \qquad r = \frac{16 n}{4 N_\star + 1} \, .
\end{equation}
Let us now consider $N_\star=60$ and take the quadratic ($n=2$) potential $U = \frac{1}{2} m^2 \phi^2$. We find
\begin{equation}
n_s \simeq  1 - \frac{2}{N_\star} \simeq 0.97 \, , \qquad r \simeq \frac{8}{N_\star} \simeq 0.13 \, .
\end{equation}
Similarly, for the quartic ($n=4$) potential $U = \frac{1}{4} \lambda \phi^4$ we find
\begin{equation} 
n_s \simeq 1 - \frac{3}{N_\star} \simeq 0.95 \, , \qquad r \simeq \frac{16}{N_\star} \simeq 0.26 \, .
\label{eq:quartic:r-ns}
\end{equation}
In both cases, $r$ is above the observational bound \eqref{eq:r} and the models are ruled out. Simple models like these are pedagogical, but more ingredients are needed to fit the observations.

\subsection{Starobinsky inflation}
\label{sec:starobinsky}

One of the oldest inflationary models is Starobinsky inflation \cite{Starobinsky:1980te}, defined by adding an $R^2$ term to the action:
\begin{equation}
S = \frac{1}{2}\int \mathrm{d}^4x\sqrt{-g} \left( R +\alpha R^2\right) \, .
\end{equation}
In metric gravity, the extra term introduces a new propagating degree of freedom, which becomes the inflaton. The action---given here in the \emph{Jordan frame}---can be reduced to the standard \emph{Einstein frame} form \eqref{eq:canonical_S} with only a standard Einstein--Hilbert term in the gravity sector by performing a Weyl rescaling of the metric $g_{\mu\nu}$.

Let us go through this procedure for a general $F(R)$ theory, where
\begin{equation} \label{eq:SF_metric}
S = \frac{1}{2}\int \mathrm{d}^4x\sqrt{-g} \, F(R) \, .
\end{equation}
Introducing an auxiliary scalar field $\chi$ (traditionally called the scalaron), this can be rewritten as 
\begin{equation} \label{eq:SF_metric_chi}
S[g_{\mu\nu},\chi] = \frac{1}{2} \int \dd^{4}x \sqrt{- g}~ \left[ F'(\chi) (R - \chi) + F(\chi) \right] \, .
\end{equation}
Variation with respect to $\chi$ gives the constraint $F''(\chi)(R-\chi) = 0 \Rightarrow \chi=R$, which reduces \eqref{eq:SF_metric_chi} back into \eqref{eq:SF_metric}, proving the equivalence of the two actions. Now, with \eqref{eq:SF_metric_chi}, we define the new, Weyl rescaled metric $\bar{g}_{\mu\nu} = F'(\chi)g_{\mu\nu}$. $R$ depends on $g_{\mu\nu}$ in a complicated way through the Levi-Civita connection and transforms as \cite{Dabrowski:2008kx} 
\begin{equation} \label{eq:R_wayl_trans_metric}
    R = F'(\chi)\left( 
\bar{R} - 6\sqrt{F'(\chi)}\bar{\nabla}^\mu\bar{\nabla}_\mu \frac{1}{\sqrt{F'(\chi)}}\right) \, .
\end{equation}
On the right-hand side, we use the new, rescaled metric in the Levi-Civita connection in $\bar{R}$ and $\bar{\nabla}_\mu$.
After some algebra and integration by parts, we get
\begin{equation} \label{eq:SF_metric_Einstein}
    S = \int \mathrm{d}^4x \sqrt{-\bar{g}} \left[\frac{1}{2}\bar{R} - \frac{3}{4}\bar{g}^{\mu\nu}\bar{\nabla}_{\mu}\ln F'(\chi)\bar{\nabla}_{\nu}\ln F'(\chi)
 - \frac{1}{2} \frac{\chi F'(\chi) - F(\chi)}{F'(\chi)^{2}}\right] \, ,
\end{equation}
which is of the standard form \eqref{eq:canonical_S} when we define the new field
$\phi$ and its potential as
\begin{equation} \label{eq:chi_to_phi}
  \phi =  \sqrt{\frac{3}{2}}\ln F'(\chi) \, , \qquad
  U = \frac{1}{2} \frac{\chi F'(\chi) - F(\chi)}{F'(\chi)^{2}} \, .
\end{equation}
For the $\left( R + \alpha R^2 \right)$ model, \eqref{eq:chi_to_phi} gives
\begin{equation}  \label{eq:starobinsky_U}
U(\phi) = \frac{1}{8 \alpha} \left[ 1- \exp\left(-\sqrt{\frac{2}{3}}\phi\right)\right]^2 \, .
\end{equation}
The potential exhibits a plateau at large $\phi$, where inflation happens. With the canonical potential at hand, the procedure of section~\ref{sec:SR_and_CMB} gives, when $N_\star = 60$,
\begin{equation} \label{eq:starobinsky_CMB}
n_s = 1 - \frac{2}{N_\star} = 0.9667 \,, \qquad r = \frac{12}{N^2_\star} = 0.0033 \, ,
\end{equation}
compatible with the CMB observations. Below, we will repeatedly use similar transformations to convert non-minimal Palatini models into a canonical form, where their CMB predictions are easy to compute. Plateau potentials with an exponentially dying $\phi$-dependence are a common consequence of this procedure, and lead---quite generically---to the same $n_s$ prediction.

\section{Scalar-tensor theories}
\label{sec:scalar_tensor}

Let us now move on to consider models with Palatini gravity. We will start with simple scalar-tensor theories, where the gravity sector only contains the standard Ricci scalar $\Ri$, but it may be multiplied by a non-trivial function of the inflaton field $\phi$. We will compare the metric and Palatini formulations of such models to give a taste of the different physics that arise.

\subsection{Equivalence of Einstein--Hilbert action in metric and Palatini gravity}
\label{sec:metric_palatini_equivalence}
As a first example of Palatini gravity, let us consider the simple Einstein--Hilbert (E--H) action
\begin{equation}
\label{eq:E-H_1}
S=\frac{1}{2}\int {\rm d}^4x \sqrt{-g}\, \mathcal{R}\,. \end{equation}
We remind the reader that $\mathcal{R}$ is the full Ricci scalar, dependent on the independent connection $\tns{\Gamma}{^\rho_\mu_\nu}$. Decomposing it with equation~\eqref{eq:ricci_C}, the E--H action takes the form
\begin{equation}
\label{eq:E-H_2}
S=\frac{1}{2}\int {\rm d}^4x \sqrt{-g} \left(R +\widetilde{\nabla}_{ \rho}\tns{C}{^\rho^\mu_\mu}-
\widetilde{\nabla}_{ \mu}\tns{C}{_\rho^\rho^\mu}+
\tns{C}{^\lambda^\mu_\mu}\tns{C}{^\rho_\rho_\lambda}
-\tns{C}{^\lambda_\rho^\mu}\tns{C}{^\rho_\mu_\lambda}\right)\, .
\end{equation}
The terms with covariant derivatives are just divergences and do not
affect the equations of motion. Varying the action~\eqref{eq:E-H_2} with respect to the  distortion tensor $\tns{C}{^\rho_\mu_\nu}$ we obtain the equation (see~\cite{Dadhich:2012htv} for an analytic derivation)
\begin{equation}
\label{eq:eom_dis}
g_{\alpha\beta}\tns{C}{^\mu_\mu_\gamma} +g_{\alpha\gamma}\tns{C}{_\beta_\mu^\mu} -C_{\alpha\beta\gamma} -C_{\beta\gamma\alpha} = 0\,.    
\end{equation}
With some algebra, we obtain that the solution of~\eqref{eq:eom_dis} is $C_{\alpha\beta\gamma}=g_{\alpha\gamma}U_\beta$, for an arbitrary vector $U_\beta$. Substituting this into \eqref{eq:E-H_2}, the last two terms cancel, and we are left with only the standard Levi-Civita Ricci term $R$.

For the pure E--H action without matter, the metric and Palatini formulations then produce the same dynamics for the metric $g_{\mu\nu}$. This remains true even if we add a matter term to action \eqref{eq:E-H_1}, as long as it does not depend on $\tns{C}{^\rho_\mu_\nu}$. We will exploit this feature below to deal with models that deviate from the E--H action \eqref{eq:E-H_1}: if we can reduce such a model to the standard form \eqref{eq:canonical_S}, it becomes equivalent to its metric counterpart, and we can use the standard results of section~\ref{sec:SR_and_CMB} to compute its CMB predictions. After inflation, when the inflaton field value $\varphi$ and the curvature scalar $\Ri$ are small, the standard E--H form is also usually regained. The non-minimal gravity terms we consider below do not affect late-Universe dynamics, apart from section~\ref{sec:quintessece} on quintessential inflation.

\subsection{Scalar-tensor gravity: metric vs. Palatini}

The Jordan frame action of a general scalar-tensor theory is
\begin{equation} \label{eq:scalar_tensor_SJ}
S = \int \dd^4x \sqrt{-g}\left(\frac{1}{2} A(\varphi) \Ri - \frac{1}{2} B(\varphi) g^{\mu\nu}\nabla_{\mu}\varphi\nabla_{\nu}\varphi - V(\varphi) \right) \, .
\end{equation}
The metric and Palatini versions of this theory differ because of the non-minimal gravitational coupling of the scalar field through $A(\varphi)$. In the Palatini version, variation with respect to $\tns{\Gamma}{^\rho_\mu_\nu}$ gives the distortion tensor as
\begin{equation}
\tns{C}{^\lambda_\alpha_\beta} =  \left[\delta^{\lambda}_{\alpha} \nabla_{\beta} \omega(\varphi) +
\delta^{\lambda}_{\beta} \nabla_{\alpha} \omega(\varphi) - g_{\alpha \beta} \nabla^{\lambda}  \omega(\varphi) \right] \, , \quad \omega\left(\varphi\right)\equiv\ln\sqrt{A(\varphi)} \, .
\end{equation} 
We explicitly see that $\tns{C}{^\lambda_\alpha_\beta}$ is non-zero. However, instead of working with the distortion tensor in the Jordan frame, we perform a Weyl transformation into the Einstein frame, in both Palatini and metric formulations: 
\begin{equation} \label{eq:scalar_tensor_weyl_trans}
	\bar{g}_{\mu\nu} \equiv A(\varphi) g_{\mu\nu} \, , \quad \sqrt{-g} \equiv A^{-2}\sqrt{-\bar{g}}\,, \quad 
\Ri =  A\left(\bar{R} - \kappa \times 6\, A^{1/2}\bar{\nabla}^\mu\bar{\nabla}_\mu A^{-1/2} \right)\,,
\end{equation}
where $\kappa=1$ in metric and $\kappa=0$ in Palatini, and $\bar{R}$ and $\bar{\nabla}_\mu$ are computed with the Einstein frame Levi-Civita connection\footnote{Technically, the Palatini case gives $\Ri = A\bar{\Ri}$; the identification $\bar{\Ri}=\bar{R}$ happens based on the results of section~\ref{sec:metric_palatini_equivalence} when we see below that the gravity sector is of the standard Einstein--Hilbert form.}.
In Palatini gravity, the Weyl transformation does not touch $\Ri_{\mu\nu}(\Gamma)$, which only depends on the independent connection; only the metric in the combination $\Ri=g^{\mu\nu}\Ri_{\mu\nu}$ gets scaled. However, in the metric case, the metric-dependent Levi-Civita connection in $\Ri_{\mu\nu}(\widetilde{\Gamma})$ also transforms, as shown in equation \eqref{eq:R_wayl_trans_metric} for  Starobinsky inflation, producing the extra term.

The action is now
\begin{equation}
S = \int \dd^4x \sqrt{-\bar{g}}\left(\frac{1}{2} \bar{R} - \frac12\left(\frac{B}{A} + \kappa\times \frac32\frac{A_{,\varphi}^2}{A^2} \right)\bar{\nabla}_{\mu}\varphi\bar{\nabla}^\mu\varphi - \frac{V}{A^2} \right) \, ,
\end{equation}
and we still need a field redefinition $\varphi = \varphi(\phi)$ to make it canonical. This is accomplished with
\begin{equation}
	\frac{\dd\varphi}{\dd\phi} = \sqrt{\frac{A^2}{A B +\kappa\times \frac32 A_{,\varphi}^2}} \,,
\label{eq:scalar-tensor:field-redefinition}
\end{equation}
so that the final action is
\begin{equation}
	S = \int \dd^4x \sqrt{-\bar{g}}\bigg(\frac{1}{2}\bar{R} -\frac{1}{2}\bar{\nabla}_{\mu}\phi\bar{\nabla}^{\mu}\phi - U(\phi)  \bigg) \,, \quad U(\phi) = \frac{V(\varphi(\phi))}{A^{2}(\varphi(\phi))} \, .
\end{equation}
The non-canonical factors $A$ and $B$ in \eqref{eq:scalar_tensor_SJ} have been fully absorbed into the new potential $U$.

The difference between the formulations ultimately arises from the $A_{,\varphi}^2$ term in the field redefinition~\eqref{eq:scalar-tensor:field-redefinition}. If $A(\varphi) = 1 + \xi \varphi^2$, where $\xi$ is the non-minimal coupling, the inflationary predictions will diverge for growing $\xi$. However, 
in the small-$\xi$ limit, when $\xi \lesssim 0.0047$ \cite{Karam:2021wzz}, the models become indistinguishable for a generic potential since the $6 \kappa \xi$ difference in~\eqref{eq:scalar-tensor:field-redefinition} will be too small to be resolved experimentally. In this limit, the observable 50--60 $e$-folds of inflation revert back to the chaotic form discussed in section~\ref{sec:chaotic_inflation}. Now, for general chaotic models described by the model functions $A(\varphi) = 1 + \xi \varphi^n$, $B(\varphi) = 1$, $V(\varphi) = \lambda_{2n} \varphi^{2n}$~\cite{Jarv:2017azx}, it is easy to show~\cite{Kallosh2014b} that for large $\varphi$ the Einstein frame potential becomes $U(\phi)_{\rm metric} \simeq \frac{\lambda_{2n}}{\xi^2} \left[ 1 - \exp\left(-2 \sqrt{\frac{\xi}{1+6\xi}}\phi \right) \right]^2$, which is the Starobinsky inflation attractor for $\xi \to \infty$, with the known predictions of $n_s = 1 - 2/N_\star$ and $r = 12 / N^2_\star$. On the other hand, in the Palatini formulation for $\xi \to \infty$ the Starobinsky attractor is lost and one finds $n_s \simeq 1 - \left( 1 + \frac{n}{2} \right) \frac{1}{N_\star}$ and $r \simeq 0$. Interestingly, in the Palatini case the attractor is lost also in the case of multifield inflation~\cite{Carrilho:2018ffi}.

It is interesting to note that, despite the differences between the two formulations, one can choose the model functions $A(\varphi)$, $B(\varphi)$, $V(\varphi)$ in such a way that the Einstein frame potential $U(\phi)$ will be the same in both metric and Palatini, thus yielding the same inflationary observables. For example, 
it was shown in \cite{Jarv:2020qqm} that $n_s$ and $r$ are the same if $A(\varphi) B(\varphi) \propto \left( A'(\varphi) \right)^2$ and $V(\varphi) \propto A(\varphi)^2 \left( \ln \frac{A(\varphi)}{A_0} \right)^2$. In the same spirit,~\cite{Racioppi:2018zoy} showed that for $A(\varphi) = 1 + \xi \varphi^2$ or $A(\varphi) = \xi \varphi^2$ and a Higgs-like potential $V(\varphi) = \frac{1}{4} \lambda(\varphi) \varphi^4$ with a running quartic coupling $\lambda(\varphi)$, both the metric and Palatini versions of the theory end up in the linear inflation attractor ($n_s = 1 - \frac{3}{2N_\star}$ and $r = \frac{4}{N_\star}$) for $\xi \gtrsim 1$.

\subsection{Example: Non-minimal Higgs inflation}
\label{Example:_Higgs_Inflation}

As an example, we consider non-minimal Higgs inflation, first discussed in the metric context by Bezrukov and Shaposhnikov~\cite{Bezrukov:2007ep} and in Palatini gravity by Bauer and Demir~\cite{Bauer:2008zj}. The model is defined by
\begin{equation}
    V(\varphi) = \frac{\lambda}{4} \varphi^4 \, , \qquad A(\varphi) = 1 + \xi \varphi^2 \, , \qquad B(\varphi) = 1 \, .
\end{equation}
The procedure of the last section gives
\begin{equation}
     \varphi(\phi) \simeq \frac{1}{\sqrt{\xi}} \exp\left(\sqrt{\frac{1}{6}} \phi\right) \quad (\rm{metric})  \, , \quad
    ~~\varphi(\phi) = \frac{1}{\sqrt{\xi}} \sinh(\sqrt{\xi} \phi) \quad ~(\rm{Palatini}) \, ,
\end{equation}
so that the Einstein-frame potential can be expressed in terms of $\phi$ as
\begin{alignat}{2} \label{eq:higgs_U_metric}
U(\phi) & \simeq  \frac{\lambda}{4 \xi^{2}}\left(1-\exp \left(-\sqrt{\frac{2}{3}} \, \phi \right)\right)^{2}
\quad&&\rm{(metric)} \, , \\
\label{eq:higgs_U_Palatini}
U(\phi) &=  \frac{\lambda}{4 \xi^{2}} \tanh ^{4}\left(\sqrt{\xi} \phi \right)
&&\rm{(Palatini)\, .} 
\end{alignat}
\begin{figure}
\centering
\includegraphics[width=0.75\textwidth]{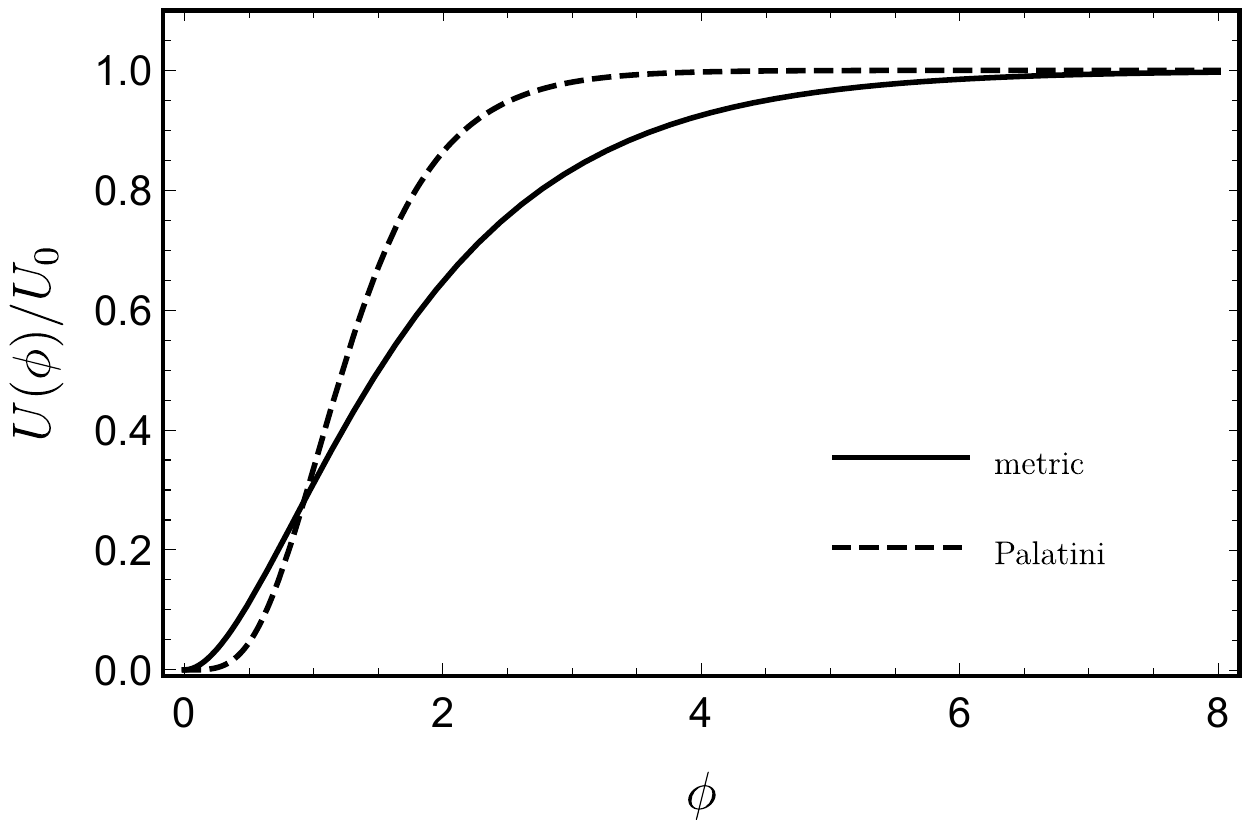}
\caption{The normalized Einstein-frame potential $U(\phi)/U_0$ with $U_0 \equiv \lam/(4 \xi^2)$, in the metric (solid curve) and Palatini (dashed curve) formulations of gravity. For illustrative purposes, we have set $\xi=1$.}
\label{fig:Higgs_MvsP}
\end{figure}
In the Palatini case, the result is exact; in the metric case, it is the leading contribution for large field values when $\xi\varphi^2 \gg 1$ and the potential becomes a plateau, see figure~\ref{fig:Higgs_MvsP}. For inflation on the plateau, the observables become
\begin{alignat}{4}
&n_{s} \simeq 1-\frac{2}{N_{*}} + \frac{3}{2 N^2_{*}}, \qquad 
&&r \simeq \frac{12}{N_\star^2}, \qquad 
&&A_s \simeq \frac{\lambda N_{*}^{2}}{72 \pi^{2} \xi^{2}} \qquad 
&&(\rm{metric}) \, , \\
&n_{s} \simeq 1-\frac{2}{N_{*}} - \frac{3}{8 \xi N^2_{*}}, \qquad
&&r \simeq \frac{2}{\xi N_{*}^{2}}, \qquad
&&A_s \simeq \frac{\lambda N_{*}^{2}}{12 \pi^{2} \xi} \qquad
&&(\rm{Palatini}) \, .
\end{alignat}
For small values of $\lambda$, $\xi$ has to be large to satisfy the constraint on $A_s$. The $n_s$ values of the models are then the same, but the Palatini model predicts a much smaller $r$, making the metric and Palatini versions observationally distinguishable. The metric model, with $r\simeq 0.0033 \dots 0.0048$ for $N_\star = 50 \dots 60$, will be probed by the next generation CMB experiments (see \textit{e.g.}~\cite{Matsumura2016, Kogut_2011, Sutin:2018onu}), but the Palatini model can easily evade any future bounds on $r$. A small $r$ turns out to be a common feature of many Palatini plateau models.

After their original paper~\cite{Bauer:2008zj}, Bauer and Demir reconsidered the Palatini version of the Higgs inflation model in the context of tree-level unitarity violation~\cite{Bauer:2010jg}\footnote{See also~\cite{Jinno:2018jei, Reyimuaji:2020goi} for other non-minimally coupled models in the Palatini formalism.}. In metric Higgs inflation, the energy scale $\Lambda$ at which tree-level unitarity is violated (at least naively, around the electroweak minimum of the potential) is of the order of $\Lambda \sim M_{\rm Pl} / \xi \sim 10^{14} \ \rm GeV $ (where we temporarily reintroduced the Planck mass), which is below the scale of inflation~\cite{Barbon2009}. As a result, it is difficult to determine the inflationary potential from the low-energy Standard Model parameters, unless the abrupt changes in the coupling constants at the beginning of the strong coupling regime are very small~\cite{Bezrukov2011, Bezrukov2015}. Moreover, the small value of $\Lambda$ is expected to result in the breakdown of perturbation theory during preheating~\cite{Ema:2016dny}. In contrast, Palatini Higgs inflation sets $\Lambda = M_{\rm Pl}/\sqrt{\xi} \sim 10^{14} \ \rm GeV $, which is above the energy scale of inflation~\cite{Bauer:2010jg}. Therefore, Palatini Higgs inflation does not suffer from tree-level unitarity violation. The issue of unitarity was further explored in~\cite{Shaposhnikov:2020geh, McDonald:2020lpz, Mikura:2021clt, Ito:2021ssc, Antoniadis:2021axu, Panda:2022esd}, and in~\cite{Ito:2021ssc} the upper limit $\xi < 2.5 \times 10^{31}$ was obtained from direct collider constraints in the Palatini formulation. The corresponding limit in the metric case is $\xi < 2.6 \times 10^{15}$~\cite{Atkins:2012yn}, derived using data on the global Higgs branching ratio from the ATLAS and CMS experiments at the LHC, with the updated limits given in~\cite{Xianyu:2013rya, Ren:2014sya, Ren:2014sya, Wu:2019hso}.

Furthermore, 
Refs.~\cite{Rasanen:2017ivk, Enckell:2018kkc, Rasanen:2018fom, Shaposhnikov:2020fdv, Enckell:2020lvn} considered loop corrections to the Higgs potential. These loop corrections can actually deform the EF potential and produce a (quasi-)inflection point. Then, if the model parameters are tuned, the inflaton can enter a short ultra-slow-roll phase around the inflection point (called critical point inflation), which can result in an amplification of the scalar power spectrum and the production of a sizeable population of primordial black holes, though they are light and evaporate quickly~\cite{Rasanen:2018fom}. Regarding the inflationary predictions of critical point Higgs inflation, in the metric case the tensor-to-scalar ratio can be enhanced from the plateau value up to $r=0.2$, while in the Palatini case $1\times 10^{-13} < r < 7 \times 10^{-5}$~\cite{Rasanen:2017ivk}. Additionally,~\cite{Jinno:2019und} studied Higgs inflation in both metric and Palatini formulations with the inclusion of higher-dimensional operators. They found that for the metric case the inflationary predictions are relatively stable against these operators because of the existence of an attractor, but in the Palatini case the higher-order operators spoil the prediction of $n_s$.

Finally,~\cite{He:2022xef} studied Higgs inflation in the context of the hybrid metric-Palatini formalism\footnote{For a review on hybrid metric-Palatini gravity see~\cite{Capozziello:2015lza}.}, where the action contains both the metric-type and the Palatini-type Ricci scalars, while~\cite{Eadkhong:2023ozb} studied the metric and Palatini versions of non-minimally-coupled warm Higgs inflation, where the inflaton is assumed to be coupled to a thermal bath and transfer energy to radiation during inflation, thus maintaining a non-zero temperature~\cite{Berera:1995wh}.

\section{Non-minimal derivative coupling}
\label{sec:non_min_derivative_coupling}
In this section, we consider the inflationary predictions for the Palatini Higgs inflation scenario, in the presence of non-minimal derivative coupling (NMDC) terms in the action\footnote{Various cosmological implications of NMDC-extended models in the Palatini formalism have also been investigated in \cite{Gumjudpai:2016ioy,Galtsov:2020jnu}. Higgs inflation with a NMDC in the metric formulation was originally proposed in~\cite{Germani:2010gm} (see also~\cite{Sheikhahmadi:2016wyz, Karydas:2021wmx}).}~\cite{Gialamas:2020vto}. Such terms were considered for the first time in~\cite{Amendola:1993uh}, as a means to further explore the influence of different kinds of non-minimal couplings in Cosmology beyond the standard non-minimal coupling of the scalar field with the Ricci scalar $A(\vphi)R$, where $A(\vphi)$ is an arbitrary smooth function of the field.

In principle, by restricting the analysis to terms that are linear in $R$, quadratic in $\vphi$, and containing four derivatives, all the possible NMDC terms that can be constructed are $\kappa_1 R \nabla_{\mu}\vphi \nabla^{\mu}\vphi$, $\kappa_2 R_{\mu\nu}\nabla^{\mu}\vphi\nabla^{\nu}\vphi$, $\kappa_3 R \vphi\square\vphi$, $\kappa_4 R_{\mu\nu} \vphi\nabla^{\mu}\nabla^{\nu}\vphi$, $\kappa_5 \nabla_{\mu}R \vphi \nabla^{\mu}\vphi$ and $\kappa_6 \square R \vphi^2$ where $\kappa_i,\, i=1,\ldots,6$ are coupling constants with dimensions of \emph{mass}$^{-2}$ \cite{Amendola:1993uh}. It has been shown that only the first two of the aforementioned terms suffice to encapsulate the properties of these theories~\cite{Capozziello:1999xt, Sushkov:2009hk} (see also \cite{Amendola:1993uh, Capozziello:1999uwa}). For arbitrary values of $\kappa_1$ and $\kappa_2$, the resultant field equations contain third-order derivatives, however, for $\kappa_2=-2\kappa_1$, the field equations are of second order~\cite{Sushkov:2009hk} thus avoiding instabilities of the theory \cite{Ostrogradsky:1850fid, Woodard:2015zca}. Abiding by this constraint, the NMDC sector of the action consists of a coupling between the Einstein tensor and derivatives of the scalar field in the form $\kappa_2 G_{\mu\nu} \nabla^{\mu}\vphi\nabla^{\nu}\vphi$. In~\cite{Dalianis:2019vit}, a generalization of this NMDC term was introduced, where the coupling constant is promoted to a function of the scalar field $\kappa_2 \to m^{-2}(\vphi)$.

In order to develop the formalism in a general way, our starting point will be the Palatini action
\begin{equation}
S = \int \dd^4x \sqrt{-g} \left[ \frac{A(\varphi)}{2} \Ri +\frac{1}{2 m^2(\varphi)}G^{\mu\nu}\nabla_\mu \varphi \nabla_\nu \varphi  -\frac{1}{2}\nabla_\mu \varphi \nabla^\mu \varphi -V(\varphi) \right]\,,
\label{eq:NMDC_JF_action}
\end{equation}
with $A(\vphi)$, $m(\vphi)$ and $V(\vphi)$ model functions to be specified, and $G^{\mu\nu} \equiv \Ri^{\mu\nu} -\frac{1}{2}g^{\mu\nu} \Ri\,.$ A Weyl rescaling of the metric which is usually applied in order to recast the action from the JF to the EF turns out to be insufficient for~\eqref{eq:NMDC_JF_action} due to the presence of the NMDC term. To this end, one has to resort to the more general class of \emph{disformal} transformations~\cite{Bekenstein:1992pj,Zumalacarregui:2013pma,Minamitsuji:2014waa,Tsujikawa:2014uza,Domenech:2015hka,Sakstein:2015jca,vandeBruck:2015tna,Achour:2016rkg,Gumjudpai:2016ioy,Escriva:2016cwl,Lobo:2017bfh,Sato:2017qau,Galtsov:2018xuc, Karwan:2018eln,Delhom:2019yeo,Galtsov:2020jnu,Qiu:2020csx}, under which the metric changes according to
\begin{equation}
g _{\mu\nu}\equiv \Omega ^2(\vphi,X) \left[\bar{g}_{\mu\nu} +\beta ^2(\vphi,X) \bar{\nabla} _\mu \vphi\, \bar{\nabla} _\nu \vphi \right] \,, \quad X\equiv-\frac{1}{2} \bar{\nabla} _\mu \vphi \bar{\nabla} ^\mu\vphi\,.
\label{eq:NMDC_disformal_def}
\end{equation}
Once the transformation is applied, one has to determine the functional dependence of $\Omega^2$ and $\beta^2$ on the field $\vphi$ and its velocity $-u^2\equiv u_{\mu}u^{\mu}=\bet^2 \left(\bar{\nabla}\vphi\right)^2$ such that the standard EF action is obtained. For~\eqref{eq:NMDC_JF_action}, the appropriate disformal factors are~\cite{Gialamas:2020vto}
\beq 
\Omega^2= \frac{2- u^2}{2 A(\vphi) \sqrt{1- u^2}}\quad \text{and} \quad \beta ^2 =\frac{1}{m^2(\vphi) \sqrt{1- u^2}}\,.
\label{omega_beta_m}
\eeq
The kinetic and potential terms in the EF action obtained after the disformal transformation contain prefactors that depend on both $\vphi$ and $X$. In order to separate the two, and to be able to compute the inflationary observables, one has to work under the SR approximation and perform an expansion around small values of $X$ (see~\cite{Gialamas:2020vto} for more details). Eventually, we find
\begin{equation}
S \simeq \int \dd^4x \sqrt{-\bar{g}} \left[ \frac{\bar{R}}{2} - K(\vphi) \frac{ (\bar{\nabla} \vphi)^2}{2} +  L(\vphi) \frac{ (\bar{\nabla} \vphi)^4}{4} - U(\vphi)\right]\,,
\label{eq:NMDC_S_E}
\end{equation}
where the action functionals are given by
\begin{equation} 
K(\vphi) \equiv \frac{1}{A(\vphi)} + \frac{U(\vphi)}{m^2(\vphi)} \,, \quad L(\vphi) \equiv \frac{1}{m^2(\vphi)} \left[ \frac{1}{A(\vphi)} + \frac{U(\vphi)}{2 m^2(\vphi)} \right] \,, \quad U(\vphi) \equiv \frac{V(\vphi)}{A^2(\vphi)}\,.
\label{eq:NMDC_action_functs}
\end{equation}

The EF action~\eqref{eq:NMDC_S_E} contains a single non-canonical degree of freedom along with a higher-order kinetic term $\propto \left(\bar{\nabla}\vphi\right)^4$ which, however, can be ignored since it has been shown that in similar models the inflaton converges exponentially fast to the slow-roll attractor~\cite{Tenkanen:2020cvw}, thus rendering the higher-order kinetic terms irrelevant during inflation. Furthermore, it is worth noting that equation~\eqref{eq:NMDC_S_E} is similar to the action in the metric scenario, albeit with the non-canonical kinetic functions $K\left( \varphi \right)$ and $L\left( \varphi \right)$ defined differently~\cite{Sato:2017qau}. The scalar field can be rendered canonical under the field redefinition\footnote{In the metric formulation, the radicand in~\eqref{eq:NMDC_dphi/dvphi} contains the extra term $+(3/2) A_\vphi^2\left(\vphi\right)$, in analogy with~\eqref{eq:scalar-tensor:field-redefinition} (see~\cite{Sato:2017qau}).}
\begin{equation}
\frac{\dd\phi}{\dd\vphi}=\frac{\sqrt{A\left(\vphi\right)+V\left(\vphi\right)/m^2\left(\vphi\right)}}{A\left(\vphi\right)}\,.
\label{eq:NMDC_dphi/dvphi}
\end{equation}
Equation~\eqref{eq:NMDC_dphi/dvphi} cannot always be integrated analytically to yield $\phi\left(\vphi\right)$, but one can utilize the chain rule and compute the inflationary observables directly in terms of the non-canonical field. The first two potential slow-roll parameters and the number of $e$-foldss, in this case, are computed via
\begin{equation}
\epsilon_U
= 
\frac{1}{2 K} 
\left(
\frac{U'}{U}
\right)^2, \quad 
\eta_U
= \frac{1}{U \sqrt{K}} \left( \frac{U'}{\sqrt{K}} \right)', \quad
N
= 
\int\! 
K\, \frac{U}{U'}\, \dd \vphi \,,\label{eq:NMDC_PSRPs}
\end{equation}
where prime corresponds to differentiation with respect to $\vphi$.

\subsection{Example: Non-minimal Higgs Inflation}

\begin{figure}
\centering
\includegraphics[width=0.7\textwidth]{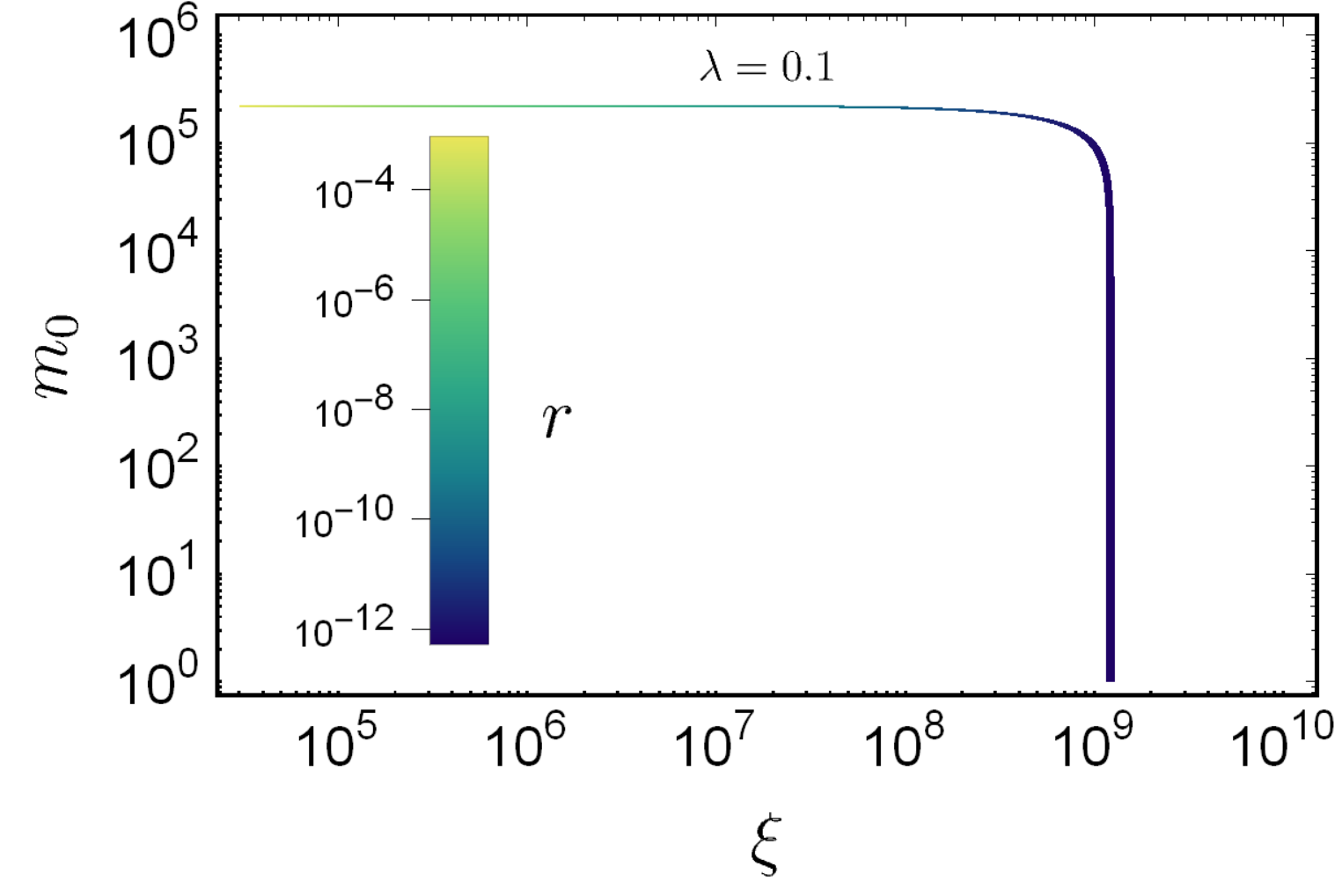}
\caption{At $N_\star=55$, for $\lambda=0.1$ 
we plot $A_s (\xi, m_0) \simeq 2.1 \times 10^{-9}$. The color grading of the curves corresponds to values of $r$ as they are depicted in the inlaid bar. For $m_0 \simeq 2.2 \times 10^5$, the factor $m_0^2 \lambda$ in the formula~\eqref{eq:NMDC_observables} for $A_s$ dominates over $\xi$. Therefore, we can have smaller values for $\xi$, which means that $r$ can be larger. For $\xi \lesssim 3 \times 10^{4}$, 
the validity of the $\sqrt{\xi} \vphi \gg 1$ approximation fails.}
\label{fig:Case1-density-r}
\end{figure}

The model functions for a quartic Higgs-like model with a quadratic coupling functional for the NMDC term\footnote{In~\cite{Gialamas:2020vto}, the authors also considered the case of a constant $m(\vphi)$ and found similar results to the ones presented here.} read
\beq
A(\vphi)=1+\xi \vphi^2\,,\quad V(\vphi)=\lambda \vphi^4/4\,,\quad m^2(\vphi) = \vphi^2/m_0^2\,,
\label{eq:NMDC_model_func}
\eeq
and the EF potential is thus given by~\eqref{eq:NMDC_action_functs}
\beq
U(\vphi)=\frac{\lam \vphi^4}{4 \left( 1+\xi \vphi^2\right)^2}\,.
\eeq
For the model functions~\eqref{eq:NMDC_model_func}, equation~\eqref{eq:NMDC_dphi/dvphi} cannot be solved analytically and so we work with the non-canonical field by means of~\eqref{eq:NMDC_PSRPs}. The integration for the number of $e$-foldss in~\eqref{eq:NMDC_PSRPs} can nevertheless be performed analytically and yields the result
\beq
N_\star=\frac{\vphi^2}{8}+\frac{m_0^2\lambda}{32\xi^2}\,\left[\xi\vphi^2-\ln{(1+\xi\vphi^2)}\right] \,.
\label{eq:CaseI_efolds}
\eeq
The first term in the above equation is the standard SR expression for the Palatini Higgs inflation model, while the second term stems from the NMDC extension of the action. In order to derive analytic expressions for the inflationary observables, we work under the approximation $\xi \vphi^2 \gg 1$, where one has
\beq
\vphi^2 \simeq \frac{32 N_\star}{4 + \frac{m_0^2 \lambda}{\xi}} \,,
\label{eq:NMDC_model_funcs}
\eeq
and so, after performing a large-$N_\star$ expansion, we obtain the compact expressions
\begin{equation}
A_s \simeq \frac{\lambda N^2_\star}{3 \pi^2 \left( m_0^2 \lambda + 4 \xi \right)} \,, \quad n_s \simeq 1 - \frac{2}{N_\star} - \frac{m_0^2 \lambda + \xi}{8 \xi^2 N_\star^2}\,, \quad r \simeq \frac{m_0^2 \lambda + 4 \xi}{2 \xi^2 N^2_\star}\,.
\label{eq:NMDC_observables}
\end{equation}
By substituting the measured value for the power spectrum $A_s \simeq 2.1 \times 10^{-9}$ in the above expression, we obtain the following constraint between the model parameters ($\xi,\lambda,m_0$) at large $N_\star$:
\beq
\xi \approx 4 \times 10^6 N_\star^2 \lambda - \frac{m_0^2 \lambda}{4} \,.
\label{eq:NMDC_As_constraint}
\eeq
In the limit $m_0 \to 0$, equations~\eqref{eq:NMDC_observables} and~\eqref{eq:NMDC_As_constraint} reduce to those of the standard Palatini Higgs inflation. The most important consequence of the constraint~\eqref{eq:NMDC_As_constraint} has to do with the fact that for fixed $N_\star$ and $\lambda$, an increase in the value of $m_0$ reduces the value of $\xi$, and this corresponds to an increase in the predicted value for $r$ according to~\eqref{eq:NMDC_observables}. This property is further illustrated in figure~\ref{fig:Case1-density-r}, where we have used a more precise analytic expression for $A_s$ with no no large-$N_\star$ expansion\footnote{This expression for $A_s$, along with the more accurate expressions for $r$ and $n_s$, can be found in appendix~A.2 of~\cite{Gialamas:2020vto}.}.

From the expressions~\eqref{eq:NMDC_observables} and a typical value for the number of $e$-foldss $N_\star=50$, we can relate the tensor-to-scalar ratio in the theory~\eqref{eq:NMDC_JF_action} to to the standard result for Palatini Higgs inflation, $r_0\equiv2/(\xi N_\star^2)$, as~\cite{Gialamas:2020vto}
\beq
r \sim r_0 \times \left(10^{10} \frac{\lambda}{\xi} \right)\,.
\eeq
For each $\xi$ and $\lambda$, the value of $m_0$ has to be adjusted accordingly in such a way that $A_s$ complies with its measured value. This feature of the NMDC-extended model provides a means for the Palatini Higgs inflation scenario (which typically predicts~$r \sim \mathcal{O}(10^{-12})$~\cite{Bauer:2008zj, Rasanen2017, Markkanen:2017tun, Takahashi2019}) to be in contact with observations in the near-future experiments that will probe the $\mathcal{O}\left(10^{-4}\right)$ regime for the tensor-to-scalar ratio~\cite{Matsumura2016,Kogut_2011,Sutin:2018onu}.

With regards to the NMDC effects on the predictions for the spectral index, the second-order correction in the large-$N_\star$ expansion for $n_s$ reads
\begin{equation}
{n_s}^{(2)}=-\frac{r}{4}+\frac{3}{8\xi N_\star^2}\simeq\left.-\frac{1}{\xi N_\star^2}\left(10^{10}\,\frac{\lambda}{\xi}-\frac{3}{8}\right)\right|_{N_\star=50},
\end{equation}
which for $\lambda=0.1$ and $\xi=10^5$ turns out to be ${n_s}^{(2)}\sim 10^{-4}$. This means that the spectral index remains essentially unaffected in this case, taking values around $n_s\sim1-2/N_\star\simeq0.96$.

\section{Palatini quadratic gravity}
\label{sec:quadratic_gravity}

This section will provide an overview of the key features of a family of Palatini theories containing an inflaton field $\varphi$ with a non-minimal coupling $A(\varphi)$, along with an $\Ri^2$ term. We also consider more general theories quadratic in the curvature tensors and finally models higher order in $\Ri$.

\subsection{Palatini \texorpdfstring{$\Ri^2$}{R\^{}2}}
\label{sec:R2}

We again begin with a Jordan frame action in the Palatini formulation, this time of the form
\begin{equation}
S = \int\dd^4 x \sqrt{-g} \left[ \frac{1}{2} A(\varphi) \Ri +  \frac{\alpha}{2} \Ri^2 - \frac{1}{2} g^{\mu\nu} \nabla_\mu \varphi \nabla_\nu \varphi - V(\varphi) \right] \,, 
\end{equation}
where $\alpha$ is assumed to be positive definite and constant\footnote{The same theory but with a field-dependent $\alpha (\varphi)$ was considered in~\cite{Lykkas:2021vax}.}.
The theory was first considered in~\cite{Meng:2004yf} and then independently in~\cite{Enckell:2018hmo} and~\cite{Antoniadis:2018ywb}. In order to bring the action to the canonical form with a minimally coupled inflation field, we eliminate the $\alpha \Ri^2$ term by introducing the auxiliary field $\chi \equiv 2\alpha \Ri$ and perform a Weyl transformation of the form
\begin{equation}
    \bar{g}_{\mu\nu} = \Omega^2 g_{\mu\nu} = [ \chi + A(\varphi) ] g_{\mu\nu} \,.
\end{equation}
Then, we obtain an action where the gravitational part takes the standard Einstein--Hilbert form
\begin{equation}
 S = \int\dd^4 x \sqrt{-\bar{g}} \left[ \frac{1}{2} \bar{R} - \frac{1}{2} \frac{1}{\chi + A(\varphi)} \bar{g}^{\mu\nu} \bar{\nabla}_\mu \varphi \bar{\nabla}_\nu \varphi - \hat V(\varphi, \chi) \right]\,, 
 \label{eq:action:PalatiniR2:Weyl}
\end{equation}
with a non-canonical kinetic term and a conformally transformed potential
\begin{equation}
\hat V(\varphi, \chi) = \frac{1}{[\chi + A(\varphi)]^2} \left[ V(\varphi) + \frac{\chi^2}{8 \alpha} \right] \,,
\end{equation}
which depends on two scalar fields: the original $\varphi$ and the auxiliary $\chi$. Note that in the action~\eqref{eq:action:PalatiniR2:Weyl}, no kinetic term has been generated for the field $\chi$. This is in contrast to the metric version of the theory, where the auxiliary field
inherits a kinetic term after the Weyl rescaling of the metric, as we saw in the case of Starobinsky inflation in section~\ref{sec:starobinsky}. Contrary to the Starobinsky case, we have to explicitly add a dynamic field $\varphi$ to play the role of inflaton. The fact that $\chi$ has a kinetic term in the metric case but not in the Palatini case is attributed to the different way the Ricci scalar transforms in the two formalisms, see equation~\eqref{eq:scalar_tensor_weyl_trans}.

Upon varying~\eqref{eq:action:PalatiniR2:Weyl} with respect to $\chi$, we obtain a constraint equation which can easily be solved to give
\begin{equation}
\chi = \frac{ 8 \alpha V(\varphi) + 2 \alpha A(\varphi) \left( \bar{\nabla} \varphi \right)^2 }{ A(\varphi) - 2 \alpha \left( \bar{\nabla} \varphi \right)^2 } \,.
\label{eq:constraint_chi}
\end{equation}
We can then eliminate $\chi$ altogether by  inserting~\eqref{eq:constraint_chi} into~\eqref{eq:action:PalatiniR2:Weyl}. We readily obtain
\begin{equation}
S = \int\dd^4 x \sqrt{-\bar{g}} \left[ \frac{1}{2} \bar{R} - \frac{1}{2} 	K(\varphi) \left( \bar{\nabla} \varphi \right)^2 + \frac{1}{4}L(\varphi)  \left( \bar{\nabla} \varphi \right)^4  - \frac{\bar{U}}{1+8\alpha \bar{U}}  \right]\,, 
\end{equation}
where we defined
\begin{equation}
    \bar{U}(\varphi) \equiv \frac{V(\varphi)}{[A(\varphi)]^2} \,,
\end{equation}
and 
\begin{equation}
K(\varphi) \equiv \frac{1}{ A (1+8\alpha \bar{U}) }\,, \qquad L(\varphi) \equiv \frac{2\alpha}{A^2 (1+8\alpha \bar{U})}\,. 
\end{equation}
It is worth noting that the potential $\bar{U}$ has the same form as the Einstein frame potential when the $\alpha \Ri^2$ term is absent. Aside from altering the potential, the conformal transformation and the elimination of the $\alpha \Ri^2$ term have generated a quartic kinetic term for $\varphi$ which could potentially become dominant for large values of $\alpha$. However, it has been demonstrated~\cite{Gialamas:2019nly, Karam:2021sno, Dimopoulos:2022rdp} that this term does not play a significant role either during the slow-roll phase or after the end of inflation.

We can make the kinetic term canonical via a field redefinition of the form
\begin{equation}
    \left(\frac{\dd \varphi}{\dd \phi}\right)^2 = A ( 1 + 8 \alpha \bar{U} ) \,.
\end{equation}
Using that, we arrive at our final action
\begin{equation}
S = \int\dd^4 x \sqrt{-\bar{g}} \left[ \frac{1}{2} \bar{R} - \frac{1}{2} \left( \bar{\nabla} \phi \right)^2 + \frac{\alpha}{2} \left( 1 + 8 \alpha \bar{U}(\phi) \right) \left( \bar{\nabla} \phi \right)^4 - U(\phi)  \right] \,, 
\label{eq:action:PalatiniR2:final}
\end{equation}
where the potential is defined as
\begin{equation}
    U \equiv \frac{\bar{U}}{1 + 8 \alpha \bar{U}}\,.
\end{equation}
This potential exhibits some interesting features:
\begin{enumerate}
    \item Regardless of the shape of the Jordan frame potential $V$, the $\alpha \Ri^2$ term decreases the height of the effective potential $U$.
    \item For large values of $\bar{U}$ (usually corresponding to large values of the field $\phi$), the potential always becomes a plateau, approaching the constant value $1/(8\alpha)$.
\end{enumerate}
We will next see how these features modify the inflationary CMB predictions.

In a flat FRW background, the Friedmann and Klein-Gordon equations obtained from the action~\eqref{eq:action:PalatiniR2:final} read
\begin{equation}
3 H^2 = \frac{1}{2} [ 1 + 3 \alpha ( 1+8\alpha \bar{U} ) \dot\phi^2 ] \dot\phi^2 + U \,,
\end{equation}
\begin{equation}
0 = [ 1+ 6 \alpha ( 1 + 8 \alpha \bar{U} ) \dot\phi^2 ] \ddot\phi + 3 [ 1 + 2 \alpha ( 1 + 8 \alpha \bar{U} ) \dot\phi^2 ] H \dot\phi + 12 \alpha^2 \dot\phi^4 \bar{U}' + U' \, .
\label{eq:PalatiniR2:Klein-Gordon}
\end{equation}
Note that for $\alpha = 0$, the above equations reduce to the usual ones~\eqref{eq:Friedmann}--\eqref{eq:klein_gordon}. 

Inflation occurs when the value of the first Hubble slow-roll parameter 
\begin{equation}
\epsilon_H \equiv -\frac{\dot{H}}{H^2} = \frac{\dot{\phi}^2}{2 H^2} \left[ 1 + 2 \alpha \left( 1 + 8 \alpha \bar{U} \right) \dot{\phi}^2 \right] 
\end{equation}
is less than one. During slow roll\footnote{The Palatini $\Ri^2$ models have also been considered in the context of constant roll\cite{Antoniadis:2020dfq, AlHallak:2021hwb, Panda:2022can}, where $\ddot{\phi} = \beta H \dot{\phi}$, with $\beta$ being the constant-roll parameter.}, $\epsilon_H \ll 1$ and the $\ddot\phi$ term in~\eqref{eq:PalatiniR2:Klein-Gordon} is negligible. Similarly, the contribution of the higher-order kinetic term in~\eqref{eq:PalatiniR2:Klein-Gordon} can be disregarded as it is subdominant during slow roll, as discussed above (see also footnote $1$ of \cite{Enckell:2018hmo}). To elaborate on the higher-order kinetic terms, their behaviour is typically described by the speed of sound parameter, which has been shown in similar Palatini-$\Ri^2$ models to remain close to unity throughout inflation and only differ towards the end, with no significant impact on inflationary predictions~\cite{Gialamas:2019nly, Karam:2021sno}. Additionally, even for large values of $\alpha$, the slow-roll attractor is reached rapidly \cite{Tenkanen:2020cvw} and does not require fine-tuning. Furthermore, the expressions for the amplitude of the scalar power spectrum $A_s$ and spectral index $n_s$ as functions of $\phi$ do not explicitly depend on $\alpha$ \cite{Enckell:2018hmo}
\begin{equation}
24 \pi^2 A_s = \frac{U}{\epsilon_U} = \frac{\bar{U}}{\epsilon_{\bar{U}}} \,, \qquad n_s = 1 - 6 \epsilon_U + 2 \eta_U = 1 - 6 \epsilon_{\bar{U}} + 2 \eta_{\bar{U}}\,.
\label{eq:PalatiniR2:As-ns}
\end{equation}
The first two potential slow-roll parameters are defined as\footnote{Prime in $\epsilon_{\bar{U}}$, $\eta_{\bar{U}}$ denotes a derivative with respect to the canonical field in a model with $\alpha=0$.}
\begin{equation} 
\epsilon_U = \frac{1}{2} \left( \frac{U'}{U} \right)^2 \,, \qquad \epsilon_{\bar{U}} = \frac{1}{2} \left( \frac{\bar{U}'}{\bar{U}} \right)^2 = \epsilon_U \vert_{\alpha = 0}\,,
\end{equation}
\begin{equation} 
\eta_U =  \frac{U''}{U} \,, \qquad \eta_{\bar{U}} =  \frac{\bar{U}''}{\bar{U}}  = \eta_U \vert_{\alpha = 0}\,.
\end{equation}
The tensor power spectrum $A_T$, in contrast, is explicitly dependent on $\alpha$,
\begin{equation} 
A_T = \frac{2}{3 \pi^2} U = \frac{2}{3\pi^2} \frac{\bar{U}}{1 + 8 \alpha \bar{U}}\,.
\end{equation}
As a result, the tensor-to-scalar ratio becomes 
\begin{equation} 
r = 16 \epsilon_U = \frac{\bar{r}}{1 + 8 \alpha \bar{U}}  = \frac{\bar{r}}{ 1+ 12 \pi^2 A_s \bar{r} \alpha } \,. \label{eq:r2}
\end{equation}
In the last expression, we utilized equation~\eqref{eq:PalatiniR2:As-ns}, while the expression $\bar{r}=16\epsilon_{\bar{U}}$ corresponds to the tensor-to-scalar ratio of the same model but without the $\alpha \Ri^2$ term. It becomes evident that by increasing $\alpha$ significantly, we can reduce the value of $r$ in a specific model, without influencing the prediction for $n_s$.

We may now establish an upper limit on the value of $\alpha$ based on the detectability of a given model in future experiments. If approved, upcoming CMB satellites such as LiteBIRD \cite{Matsumura2016}, PIXIE \cite{Kogut_2011}, and PICO \cite{Sutin:2018onu} will have the ability to detect B-mode polarization in the primordial CMB for tensor-to-scalar ratio values of $r < 0.001$. The sensitivity of PICO, in particular, is estimated to be approximately $\delta_r \approx 10^{-4}$. We can use this value to set an upper limit on $\alpha$. When $\alpha \to \infty$ (and $\bar{r} > \delta_r$), we obtain
\begin{equation}
     r_\text{limit} \approx \frac{1}{ 12 \pi^2 A_s \alpha } \,.
     \label{eq:rlimit}
\end{equation}
This means that if $\alpha < 4 \times 10^{10}$, a given model will predict a value of $r$ that could be detected by PICO. If $\alpha > 4 \times 10^{10}$, however, even though the model could still be viable, the value of $r$ will be too small to be detected in the near future.

Finally, in the large $\alpha$ limit and using equation~{\eqref{eq:e-folds:simplified}}, one can show~\cite{Gialamas:2020snr} that the number of $e$-folds is given by
\begin{equation}
    N_{\alpha \ggg 1} \simeq 60.4 - \frac{1}{4} \ln \alpha \,.
\end{equation}
Therefore, for growing $\alpha$ we will have less $e$-folds of inflation.

\subsubsection{Minimal quadratic model}
\label{sec:minimal_quadratic}

As an application of the above results, let us consider a simple example of a quadratic potential $V(\varphi) = \frac{1}{2} m^2 \varphi^2$ with a minimal coupling $A(\phi) = 1$ to gravity
, plus the $\alpha \Ri^2$ term in the Palatini formalism.

\begin{figure}
\centering
\includegraphics[width=0.75\textwidth]{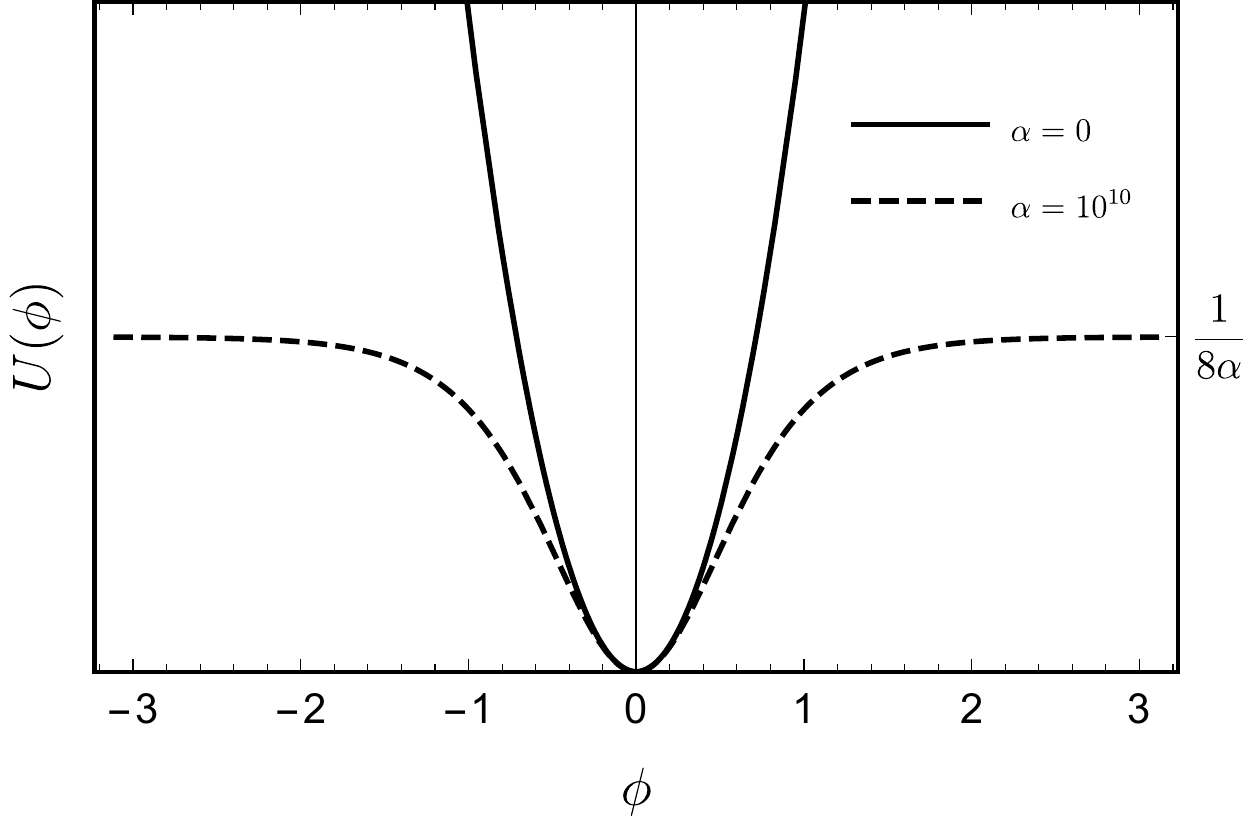}
\caption{The potential~\eqref{eq:R2_quadratic_U} for the minimal quadratic model plus the $\alpha \Ri^2$ term in the Palatini formalism for $m=7\times 10^{-6}$. The dashed line corresponds to $\alpha = 10^{10}$, while the solid line is for $\alpha = 0$, thus reducing to the minimal model with the $\alpha \Ri^2$ term absent.}
\label{fig:quad_pot}
\end{figure}

The appealing feature of this simple model is that the field redefinition can be solved and inverted analytically:
\begin{equation} \label{eq:R2_phi_transformation}
\phi = \phi_0 \sinh^{-1} \left( \frac{\varphi}{\phi_0} \right) \,, \qquad \phi_0 \equiv \frac{1}{2 m \sqrt{\alpha}} \, .
\end{equation}
Then, the potential becomes 
\begin{equation} \label{eq:R2_quadratic_U}
    U(\phi) = \frac{1}{8 \alpha} \tanh^2 \left( \frac{\phi}{\phi_0} \right) \, ,
\end{equation}
which exhibits plateaus on either side of the global minimum.
The potential is depicted in figure~\ref{fig:quad_pot}. One can see that for $\alpha = 0$ the potential reduces to a simple quadratic one, while for large values of $\alpha$ the potential develops plateaus which asymptote to $U = 1/(8\alpha)$. Even higher values of $\alpha$ would reduce the height of the potential and thus the scale of inflation. For inflation to take place above the $\rm TeV$ scale, we would need $U^{1/4} \gtrsim 10^{-15}$ in Planck units, or $\alpha \lesssim 10^{60}$.

\begin{figure}
\centering
\includegraphics[width=0.75\textwidth]{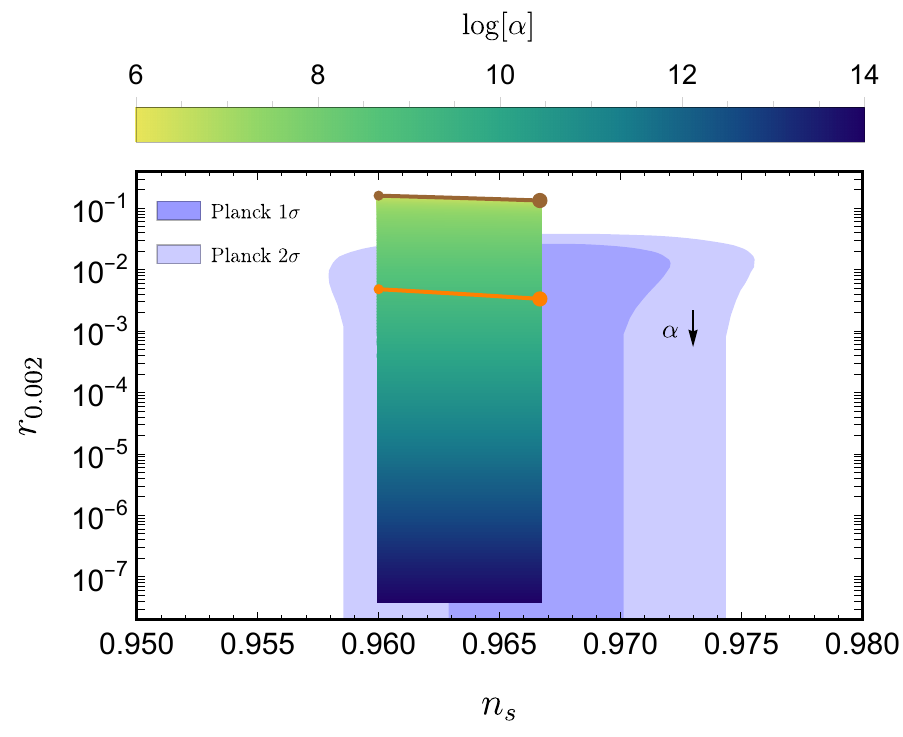}
\caption{Inflationary predictions for the minimal quadratic model in the $n_s$ vs. $r$ plane. For $\alpha=0$ we recover the prediction $r = 0.13$ of the model (brown line) without the $\alpha \Ri^2$, while for growing $\alpha$ (as indicated by the black arrow) the value of $r$ is reduced significantly. The boundaries of the shaded region correspond to $N_\star = 50$ (left) and $N_\star = 60$ $e$-folds (right). The orange line indicates the prediction of the Starobinksy model. Finally, the mass parameter $m$ is fixed by the amplitude $A_s$.}
\label{fig:nsrQuad}
\end{figure}

In the slow-roll approximation, the number of $e$-folds is computed to be
\begin{equation}
    N_\star \approx \int^\phi_0 \frac{\dd \phi_\star}{\sqrt{2 \epsilon_U}} \approx \frac{\phi^2_0}{4} \sinh^2 \frac{\phi_\star}{\phi_0} \qquad \Rightarrow \qquad \phi_\star \approx \frac{\phi_0}{2} \ln \frac{16 N_\star}{\phi^2_0} \,.
\end{equation}
The inflationary observables can be readily computed and they are given as
\begin{equation} 
    n_s \simeq  1 - \frac{2}{N_\star} 
    \,, \qquad
    r 
   \simeq \frac{8}{N_\star + 16 \alpha m^2 N^2_\star} 
    \,, \qquad
    A_s \simeq  \frac{m^2 N^2_\star}{6 \pi^2} \, .
\end{equation}
We see explicitly that $A_s$ and $n_s$ (which have the standard form) do not depend on $\alpha$, but $r$ is suppressed by it (see figure~\ref{fig:nsrQuad}). The measured value of the amplitude $A_s = 2.1 \times 10^9$ fixes
\begin{equation}
\label{eq:m_quad}
    m \simeq 7 \times 10^{-6} \,,
\end{equation}
and thus the only free parameter of the model is $\alpha$. From the observational upper bound on the tensor-to-scalar ratio $r < 0.036$, we find 
\begin{equation}
\label{eq:al_quad}
    \alpha \gtrsim 10^8 \,.
\end{equation}

It is interesting to note that the action~\eqref{eq:action:PalatiniR2:final} for the scalar field $\phi$ contains a quartic kinetic term and a non-polynomial potential, which render it non-renormalizable. This poses issues such as tree-level unitarity violation in high-energy scattering. The quartic kinetic term leads to perturbative problems at an energy scale of $\alpha^{-1/4}$, whereas the potential $U$ is relevant at the energy scale of $\phi_0$. To ensure the good behaviour of the theory, we demand
\begin{equation} \label{eq:alpha_condition}
    \alpha^{-1/4} \lesssim \phi_0 \, \qquad \iff \qquad \alpha \lesssim m^{-4} \sim 10^{20} \, .
\end{equation}
This upper bound on $\alpha$ leads to a lower bound for the tensor-to-scalar ratio $r > 4 \times 10^{-14}$. Therefore, there is a contradiction with proposals that rely on Palatini $\Ri^2$ inflation~\cite{Tenkanen:2019wsd, Tenkanen:2020cvw} for accommodating the ``Trans-Planckian Censorship" conjecture which asserts that $r \lesssim 10^{-30}$~\cite{Bedroya:2019tba} and thus requiring $\alpha \gtrsim 10^{37}$.

The minimal quadratic model we presented above was first considered in~\cite{Antoniadis:2018ywb}. In \cite{Gialamas:2019nly}, the parameter space was discussed in more detail and constraints were applied from the inflationary observables. Then, in~\cite{Lloyd-Stubbs:2020pvx}, it was pointed out that the inflaton remains sub-Planckian throughout inflation if $\alpha \gtrsim 10^{12}$, contrary to what happens in the version of the model without the $\alpha \Ri^2$ term, where the inflaton is super-Planckian during inflation. In~\cite{Gialamas:2020snr}, the potential~\eqref{eq:R2_quadratic_U} was shown to arise from different types of Coleman-Weinberg inflaton potentials, while in~\cite{Karam:2021sno} the model was studied again with a focus on tachyonic preheating (see also the discussion in section~\ref{sec:preheating}).

The next simplest model considered in the context of Palatini $\Ri^2$ models is the minimal quartic inflationary model, first studied in~\cite{Antoniadis:2018yfq} and then in~\cite{Tenkanen:2019jiq, Gialamas:2019nly}. Even though the effect of the $\alpha \Ri^2$ term is to reduce the tensor-to-scalar ratio $r$ from the value $0.26$ (c.f.~\eqref{eq:quartic:r-ns}), the value of $n_s = 0.95$ is not modified and thus the model remains excluded.

Another simple model, that of minimal natural inflation with $U(\varphi) = \Lambda^4 \left[ 1 + \cos \left( \varphi / f \right) \right]$, was studied in~\cite{Antoniadis:2018ywb} and it was shown that the addition of the $\alpha \Ri^2$ in the Palatini formalism can render the (previously excluded) model viable again.

\subsubsection{Non-minimal Higgs \texorpdfstring{$+$ $\mathcal{R}^2$}{+R\^{}2}}

A more complicated example, the mixed Higgs-$\mathcal{R}^2$ model is described by the action~\cite{Antoniadis:2018ywb,  Gialamas:2019nly}
\begin{equation}
\mathcal{S}=\int {\rm d}^4 x \sqrt{-g} \left(\frac{\mathcal{R}}{2}\left(1+\xi \varphi^2 \right) + \frac{\alpha}{2}\mathcal{R}^2 -\frac12 g^{\mu\nu}\nabla_\mu \varphi \nabla_\nu \varphi - \frac{\lambda}{4}\varphi^4\right)\,,
\end{equation}
where $\varphi$ is the Higgs field, playing the role of the inflaton.
Following the same steps as before, the action can be written as
\begin{equation}
\mathcal{S}=\int {\rm d}^4 x \sqrt{-\bar{g}} \left(\frac{\bar{R}}{2} -\frac{K(\varphi)}{2} (\bar{\nabla} \varphi)^2 + \frac{L(\varphi)}{4}(\bar{\nabla} \varphi)^4 -U(\varphi)\right)\,,
\end{equation}
with
\begin{equation}
\label{eq:K_L_mixed_Higgs}
K(\varphi) = \frac{1+\xi \varphi^2}{(1+\xi \varphi^2)^2 +2\alpha\lambda \varphi^4}\,, \quad L(\varphi) = \frac{2\alpha}{(1+\xi \varphi^2)^2 +2\alpha\lambda \varphi^4}\,,
\end{equation}
and
\begin{equation}
\label{eq:U_mixed_Higgs}
 U(\varphi) = \frac{1}{8\alpha} \frac{2\alpha \lambda \varphi^4}{(1+\xi \varphi^2)^2 + 2\alpha \lambda \varphi^4}\,.
\end{equation}
Since the kinetic term cannot be canonicalized analytically, we will proceed using the original field $\varphi$.
For $\xi \varphi^2 \gg 1$, the effective potential approaches a plateau,
\begin{equation}
U(\varphi)\left|_{\xi \varphi^2 \gg 1} \right. \simeq \frac{1}{8\left( \xi^2/(2\lambda) +\alpha \right)}\,.
\end{equation}
The function $K(\varphi)$ is also controlled by the quantity $\xi^2/(2\lambda) +\alpha$ in the large field regime $(\xi \varphi^2 \gg 1)$, which means that for $\xi^2/(2\lambda) \gg \alpha$ the effect of the $\mathcal{R}^2$ term is negligible, in contrast to the case $\alpha \gg\xi^2/(2\lambda) $.
The potential slow-roll parameters $\epsilon_U$ and $\eta_U$ (defined in~\eqref{eq:NMDC_PSRPs} for non-canonical fields) are given by
\begin{equation}
\label{eq:eps_eta_HR2}
\epsilon_U = \frac{8(1+\xi \varphi^2)}{\varphi^2\left(1+2\xi \varphi^2 +(\xi^2 +2\alpha\lambda)\varphi^4 \right)}\,, \quad \eta_U = \frac{4}{\varphi^2} \left( \frac{6(1+\xi \varphi^2)}{(1+\xi \varphi^2)^2+2\alpha\lambda \varphi^4} - \frac{3+2\xi \varphi^2}{1+\xi \varphi^2} \right)\,.
\end{equation}
In order to estimate the inflationary observables, we need to express the field value at horizon crossing as a function of the number of $e$-folds. Integrating~\eqref{eq:N_SR} we obtain $\varphi_\star = \sqrt{8N_\star +\varphi_{\rm end}^2} \simeq \sqrt{8N_\star}$, where we have dropped the contribution from the field value at the end of inflation, since\footnote{Solving the equation $\epsilon_U(\varphi_{\rm end}) =1$ we obtain that $2\alpha\lambda  \varphi_{\rm end}^6 +(1+\xi  \varphi_{\rm end}^2)( \varphi_{\rm end}^2(1+\xi  \varphi_{\rm end}^2)-8)= 0$ which is a cubic equation in $ \varphi_{\rm end}^2$ so analytic solutions can be obtained. It is easily proven~\cite{Gialamas:2019nly} that there is only one real positive solution for $ \varphi_{\rm end}^2$ that is lower that 8.} $ \varphi_{\rm end} < \sqrt{8}$. The inflationary observables are given by
\begin{equation}
n_s = 1 - \frac{2}{N_\star} -\frac{1}{N_\star(1+8\xi N_\star)}\,,\quad r = \frac{2}{\xi N_\star^2}\left( \frac{\xi^2}{\xi^2 + 2\alpha\lambda} \right) \,,
\end{equation}
and
\begin{equation}
\label{eq:As_HR2}
A_s = \frac{2\lambda}{3\pi^2} \frac{N_\star^3}{1+8\xi N_\star}\,.
\end{equation}
From~\eqref{eq:As_HR2} and the observed value $A_s = 2.1\times 10^{-9}$, we obtain that the quartic and the nonminimal couplings must satisfy
\begin{equation}
\label{eq:laxiAs}
\lambda \simeq 3\times 10^{-8} \frac{1+8\xi N_\star}{N_\star^3}\,.
\end{equation}
Substituting $\xi \sim 10^{\nu}$ into~\eqref{eq:laxiAs} we obtain that $\xi^2/(2\lambda) \sim 10^{10+\nu}$, so in this case the effect of the $\alpha \mathcal{R}^2$ term becomes noticeable for $\alpha> \xi^2/(2\lambda) \sim 10^{10+\nu}$ (see~\cite{Gialamas:2019nly} for more details).

As is evident, the spectral index increases as the parameter $\xi$ grows. Therefore for values of $\xi \gtrsim 0.1$ the spectral index is compatible with the observational data for a wide range of numbers of $e$-folds. Much smaller values of $\xi$ are not enough to satisfy the observational constraints in the usual range $50<N_\star<60$.
Finally, the observational constraint $r<r_{\rm obs}=0.036$ can again be used to establish a lower limit for the parameter $\alpha$, which is
\begin{equation}
\label{eq:a_HR2}
\alpha \gtrsim \frac{10^8 \xi N_\star \left(1-r_{\rm obs}\xi N_\star^2/2 \right)}{3r_{\rm obs}(1+8\xi N_\star)}\,.
\end{equation}
The existence of a positive lower bound on $\alpha$ requires that $\xi<2/(r_{\rm obs}N_\star^2)$. This condition does not coincide with the claim  $\xi \gtrsim 0.1$ we made previously (at least for the current observational bound on the tensor-to-scalar ratio). As a consequence, the RHS of~\eqref{eq:a_HR2} is negative and the parameter $\alpha$ is unbounded, in the sense that it can acquire any positive value.

The non-minimal Higgs $+$ $\alpha \Ri^2$ in the Palatini formalism was originally studied in~\cite{Antoniadis:2018ywb} and then in \cite{Gialamas:2019nly} and \cite{Tenkanen:2020cvw} in more detail (see also~\cite{Antoniadis:2021axu} where the unitarity violation issue is discussed). Other models with a non-minimal coupling and an $\alpha \Ri^2$ term in Palatini found in the literature include 
\begin{itemize}
    \item induced gravity with $A(\varphi) = \xi \varphi^2$, $V(\varphi) = \frac{\lambda}{4} \left( \varphi^2 - v^2 \right)^2$ \cite{Antoniadis:2018ywb}
    \item non-minimal Coleman-Weinberg inflation with $A(\varphi) = \xi \varphi^2$, $V(\varphi)= \Lambda ^4 \left\{ 1 + \left[ 4 \ln \left(\frac{\varphi}{v}\right) -1 \right] \frac{\varphi^4}{v^4} \right\}$ and $V(\varphi)=\frac{1}{8}  \beta'  \varphi ^4 \ln ^2\left(\frac{\varphi }{v}\right)$ \cite{Antoniadis:2018ywb, Gialamas:2020snr}
    \item non-minimal natural inflation with $A(\varphi) = 1 + \xi\varphi^2$, $V(\varphi) = \Lambda^4 \left[ 1 + \cos\left( \varphi / f \right) \right]$ \cite{AlHallak:2022gbv}.
\end{itemize}
The main conclusion from all these studies is that for growing $\alpha$, the value of $r$ can be in agreement with the experimental constraints.

\subsection{Palatini \texorpdfstring{$\Ri^2 + \mathcal{R}_{\mu\nu} \mathcal{R}^{\mu\nu}$}{R\^{}2 + R\^{}munu R\_munu}}

In general, besides the $\mathcal{R}^2$ term, one can add to the gravitational Lagrangian 15 more quadratic terms\footnote{The most general action that contains up to quadratic terms in Riemann curvature is \begin{eqnarray}
\label{eq:metaffqua}
\mathcal{S} &=& \int \dd^4 x \sqrt{-g} \left[  \alpha \mathcal{R}^2 + \beta_1 \mathcal{R}_{\mu\nu} \mathcal{R}^{\mu\nu} + \beta_2 \mathcal{R}_{\mu\nu} \mathcal{R}^{\nu\mu} + \beta_3 \mathcal{R}_{\mu\nu} {\widehat{\mathcal{R}}}^{\mu\nu} + \beta_4 \mathcal{R}_{\mu\nu} {\widehat{\mathcal{R}}}^{\nu\mu}  +  \beta_5 {\widehat{\mathcal{R}}}_{\mu\nu} {\widehat{\mathcal{R}}}^{\mu\nu} \right. \nonumber \\
  && \left.  + \beta_6 {\widehat{\mathcal{R}}}_{\mu\nu} {\widehat{\mathcal{R}}}^{\nu\mu}  + \beta_7 {\widehat{\mathcal{R}}}_{\mu\nu} \mathcal{R}'^{\mu\nu}  + \beta_8 \mathcal{R}'_{\mu\nu} \mathcal{R}'^{\mu\nu}  + \beta_9 \mathcal{R}_{\mu\nu} \mathcal{R}'^{\mu\nu}  + \gamma_1 \mathcal{R}_{\mu\nu\sigma\lambda} \mathcal{R}^{\mu\nu\sigma\lambda} + \gamma_2 \mathcal{R}_{\mu\nu\sigma\lambda} \mathcal{R}^{\mu\sigma\nu\lambda} \right. \nonumber  \\ 
  && \left.  + \gamma_3 \mathcal{R}_{\mu\nu\sigma\lambda} \mathcal{R}^{\nu\mu\sigma\lambda} + \gamma_4 \mathcal{R}_{\mu\nu\sigma\lambda} \mathcal{R}^{\nu\sigma\mu\lambda}  + \gamma_5 \mathcal{R}_{\mu\nu\sigma\lambda} \mathcal{R}^{\sigma\nu\mu\lambda} + \gamma_6 \mathcal{R}_{\mu\nu\sigma\lambda} \mathcal{R}^{\sigma\lambda\mu\nu}  \right]\,.  
\end{eqnarray}  } that are constructed through the Riemann tensor and the three 2-rank tensors (``Ricci tensors") introduced in~\eqref{eq:Ricci_tensor}. The majority of these terms suffer from ghost degrees of freedom~\cite{BeltranJimenez:2019acz, BeltranJimenez:2020sqf, Annala:2022gtl}\footnote{The authors of~\cite{Borunda:2008kf} compare the Einstein equations in metric and Palatini formulations for actions that contain some of the Riemann squared contractions involved in the full quadratic action~\eqref{eq:metaffqua}. They found that for a certain class of theories, \textit{e.g.} in Lovelock theories of gravity~\cite{Lovelock:1971yv}, both metric and Palatini formulations lead to equivalent equations of motion.}. In particular, nonsymmetric Ricci tensors contain new degrees of freedom and can lead to instabilities~\cite{Damour1993a}.  A ``healthy" term to be added is $\mathcal{R}_{(\mu\nu)} \mathcal{R}^{(\mu\nu)}$, if we also assume a symmetric connection. The effect of such a term on inflationary phenomenology has been studied extensively in~\cite{Annala:2020cqj} (see also~\cite{Ghilencea:2020piz, Ghilencea:2020rxc, Gialamas:2021enw, Annala:2021zdt} for applications on Palatini quadratic gravity). We will not carry out a deeper analysis of these issues in this review, but will only present the final results in the case where the gravitational Lagrangian contains the $\mathcal{R}^2$ and the $\mathcal{R}_{(\mu\nu)} \mathcal{R}^{(\mu\nu)}$ terms.
Our starting point is the action
\begin{equation}
\label{eq:agr_act}
\mathcal{S}=\int {\rm d}^4 x \sqrt{-g} \left(\frac{\mathcal{R}}{2} +  \frac{\alpha}{2}\mathcal{R}^2 + \frac{\beta}{2} \mathcal{R}_{(\mu\nu)} \mathcal{R}^{(\mu\nu)} -\frac12 g^{\mu\nu}\nabla_\mu \varphi \nabla_\nu \varphi - V(\varphi)\right)\,.
\end{equation}
We will briefly discuss the recipe one must follow in order to get to the Einstein frame (see~\cite{Annala:2020cqj} for the complete procedure).

\textit{Step 1:} Introduce an auxiliary field $\Sigma_{\mu\nu}$ which on-shell is equal to $\mathcal{R}_{(\mu\nu)}$.

\textit{Step 2:} Introduce the new metric $\sqrt{-\bar{g}} \bar{g}^{\mu\nu}\equiv \sqrt{-g} \partial \mathcal{L}_G/\partial \Sigma_{\mu\nu}$, with $\mathcal{L}_G$ being the pure gravitational part of the Lagrangian.

\textit{Step 3:} Vary the action with respect to $g_{\mu\nu}$ and solve the metric $g_{\mu\nu}$ in terms of the new metric $\bar{g}^{\mu\nu}$ and the scalar field by means of a disformal transformation $g_{\mu\nu}=A\, \bar{g}_{\mu\nu} +B\, \nabla_\mu \varphi \nabla_\nu \varphi$.

\textit{Step 4:} Assume the slow-roll approximation in order to find the analytic form of the coefficients $A$ and $B$ and substitute the solution for $g_{\mu\nu}$ back to the action of \textit{Step 2}.

The resulting Einstein frame action reads
\begin{equation} 
\mathcal{S} =  \int \dd^4 x \sqrt{-\bar{g}}  \left[ \frac{1}{2}   \bar{R} -\frac{1}{2} \frac{(\bar{\nabla} \varphi)^2}{1+\tilde{\alpha}\, V(\varphi)} -\frac{ V(\varphi)}{1+ \tilde{\alpha} V(\varphi)}  +\mathcal{O}\left((\bar{\nabla} \varphi)^4\right)\right]\,, 
\label{eq:ACT_Rmn_EF}
\end{equation}
where we have defined $\tilde{\alpha}=8\alpha+2\beta$. The Einstein-frame action is of the same form as the one of the pure $\mathcal{R}^2$ model in section~\ref{sec:R2}. The only difference is that action~\eqref{eq:action:PalatiniR2:final} contains a second-order kinetic term while \eqref{eq:ACT_Rmn_EF} yields a series of higher-order kinetic terms instead. As before, these terms are negligible during slow-roll.

\subsubsection{Scale-invariant quadratic gravity}

In this section, we additionally study the fully scale-invariant model described by the action~\cite{Gialamas:2021enw}
\begin{eqnarray}
\mathcal{S} &=& 
\label{eq:JF_FD_action_in}
\int \dd^4 x \sqrt{-g} \left\lbrace \frac{1}{2} \left[ \left( \xi_\varphi \varphi^2 + \xi_h h^2 \right) \mathcal{R} + \alpha \mathcal{R}^2 + \beta \mathcal{R}_{(\mu\nu)} \mathcal{R}^{(\mu\nu)} \right] \right. \nonumber \\
&& \qquad \qquad \left. -\frac{1}{2} g^{\mu\nu} \nabla_\mu \varphi \nabla_\nu \varphi - \frac{1}{2} g^{\mu\nu} \nabla_\mu h \nabla_\nu h - V^{(0)}(\varphi,h)\right\rbrace\,,
\label{eq:act_agr}
\end{eqnarray}
where the classically scale-invariant potential is given by
\begin{equation}
\label{eq:pot_sc}
 V^{(0)}(\varphi,h) =\frac{1}{4}\left( \lambda_\varphi \varphi^4- \lambda_{h\varphi} h^2 \varphi^2 + \lambda_{h} h^4 \right) \,,
\end{equation}
and $h$ is the Higgs field. The matter sector contains also the SM Lagrangian with no Higgs mass term\footnote{The Higgs mass is dynamically generated through the interaction term $- \lambda_{h\varphi} h^2 \varphi^2$ included in the scale-invariant potential~\eqref{eq:pot_sc}.}  and three right-handed neutrinos. In this model, the Planck scale is dynamically generated once the scalar fields $\varphi$ and $h$ get their vacuum expectation values (VEVs), \textit{i.e.}
\begin{equation}
\label{eq:vevs_planck}
M^2_{\rm Pl} = \xi_\varphi v_\varphi^2 + \xi_h v_h^2\,.
\end{equation}
After a Weyl rescaling of the metric\footnote{Note that we have used the same symbol ($g_{\mu\nu}$) for the metric tensors in the actions~\eqref{eq:JF_FD_action_in} and~\eqref{eq:JF_FD_action}. The usual symbol $\bar{g}_{\mu\nu}$ is reintroduced in the EF action~\eqref{eq:final_action}. } and following the Gildener--Weinberg approach\footnote{The Gildener--Weinberg approach~\cite{Gildener1976a} is used to analyze models with multiple scalar fields that exhibit classical scale invariance. This approach involves identifying the flat directions of the potential in the field space, \textit{i.e.} the directions where its first derivatives with respect to each field vanish. The flatness of the potential at tree level suggests that the system's dynamics are governed by one-loop corrections that dominate along the flat directions. Through perturbative minimization at a particular energy scale, the flatness is removed by the loop corrections, and the physical vacuum of the theory is singled out from the degenerate minima along the flat directions.}, we obtain
\begin{figure}
\centering
\includegraphics[width=0.75\textwidth]{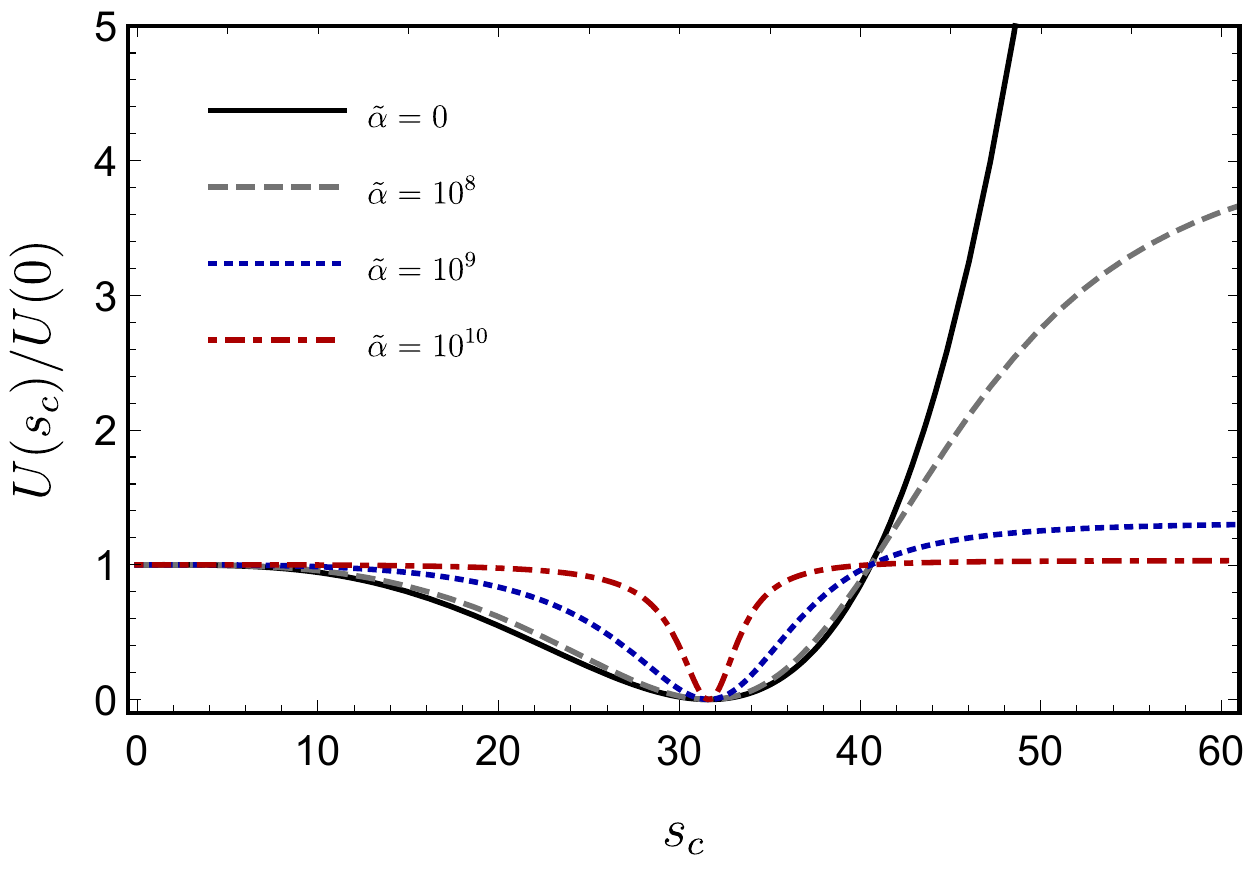}
\caption{The normalized inflationary potential given by equation~\eqref{eq:KUbar}, for various values of the parameter $\tilde{\alpha}$ as indicated in the plot. The rest parameters are fixed to $\xi_s=10^{-3}$ and $\mathcal{M}^4 = 4\times 10^{-6}$. As $\tilde{\alpha}$ gets larger values, the potential becomes symmetric about its VEV $(v_s)$ and consequently, inflation that starts from field values $s_c>v_s$ gives the same predictions with the one that starts from field values $s_c<v_s$.}
\label{fig:V_agra}
\end{figure}
\begin{equation} 
\label{eq:JF_FD_action}
\mathcal{S} = \int{\dd^4 x \sqrt{-g}}\left\lbrace \frac{1}{2} \left[ \Ri + \alpha \Ri^2 + \beta \Ri_{(\mu\nu)} \Ri^{(\mu\nu)}  \right]-\frac{1}{2} g^{\mu\nu} \nabla_{\mu} s_c \nabla_{\nu}s_c - U_{\rm eff}(s_c) \right\rbrace\,,
\end{equation}
where $s_c$ is the canonically normalized scalar field (scalon) which ``survives" along the flat direction of the initial two-field tree-level potential. The full one-loop effective potential is given by~\cite{Gialamas:2021enw}
\begin{figure}
\centering
\includegraphics[width=0.75\textwidth]{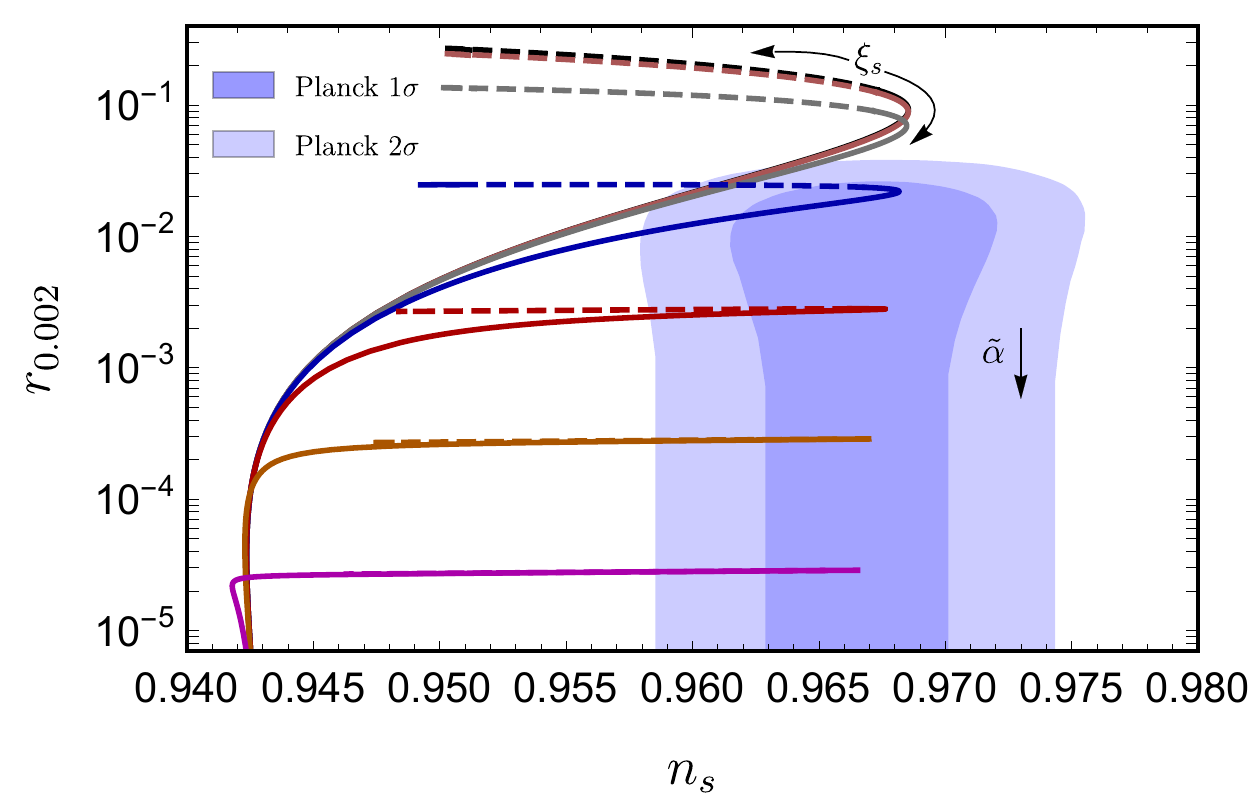}
\caption{The inflationary predictions of the model given by equation~\eqref{eq:final_action}  compared with the observational $68\%$ and $95\%$ C.L. constraints in the $n_s-r$ plane. Dashed (solid) curves  correspond to inflation that starts from field values $s_c>v_s$ ($s_c<v_s$). Same colors indicate same values for the parameter $\tilde{\alpha}$, namely $\tilde{\alpha} =0, 10^7,\cdots10^{12}$. Along the dashed (solid) curves, the parameter $\xi_s$ lies in the range $0-0.1$ ($0-10^8$). Both $\tilde{\alpha}$ and $\xi_s$ increase as indicated by the black arrows. }
\label{fig:ns_r_agra}
\end{figure}
\begin{equation}
\label{eq:V_full}
U_{\rm eff} (s_c) = \frac{\mathcal{M}^4}{128 \pi^2} \left[ \frac{s^4_c}{v^4_s} \left( 2 \ln \frac{s^2_c}{v^2_s} - 1 \right)  + 1  \right]\,,
\end{equation}
where $v^2_s\equiv M^2_{\rm Pl}/\xi_s$ is the VEV of the scalon and $\xi_s \equiv \xi_\varphi \cos^2 \omega + \xi_h \sin^2 \omega$ is an effective nonminimal coupling that is constructed from the initial nonminimal couplings and a mixing angle $\omega$.
Finally, the parameter $\mathcal{M}$ contains the masses of the particle spectrum of the  model.

The higher-order curvature terms can be eliminated by following the recipe mentioned below equation~\eqref{eq:agr_act}. The resulting EF action is~\cite{Gialamas:2021enw}
\begin{equation}
S \simeq  \int \dd^4 x \sqrt{-\bar{g}} \left[ \frac{1}{2} \bar{R} -\frac{ K(s_c)}{2} (\bar{\nabla} s_c)^2 -U(s_c)  +\mathcal{O}\left((\bar{\nabla} s_c)^4\right)\right]\,, 
\label{eq:final_action}
\end{equation}
with 
\begin{eqnarray}
K(s_c) = \frac{1}{1+\tilde{\alpha}\, U_{\rm eff}(s_c)} \quad \text{and} \quad U(s_c)= \frac{U_{\rm eff}(s_c)}{1+\tilde{\alpha}\, U_{\rm eff}(s_c)}\,.
\label{eq:KUbar}
\end{eqnarray}
In figure~\ref{fig:V_agra} we display the shape of the inflationary potential~\eqref{eq:KUbar}. As $\tilde{\alpha}=8\alpha+2\beta$ gets larger values, the potential becomes symmetric about its VEV $(v_s)$ and consequently inflation that starts from field values $s_c>v_s$  gives the same predictions with the one that starts from field values $s_c<v_s$.

The inflationary predictions of this model are shown in figure~\ref{fig:ns_r_agra} and, as expected, the usual behaviour of a decreasing scalar-to-tensor ratio is observed. In figure~\ref{fig:ns_r_agra}, the dashed parts of the curves correspond to inflation that starts from field values $s_c>v_s$, while the solid ones correspond to field values $s_c<v_s$. The two cases coincide in the limit $\xi_s \rightarrow 0$. As we move away from $\xi_s \rightarrow 0$, $\xi_s$ increases up to $\sim10^8$ ($10^{-1}$) in the dashed (solid) curves. Regarding the parameter $\tilde{\alpha}$, the upper black curve corresponds to the case $\tilde{\alpha}=0$, while for the rest of the curves it varies from $10^7$--$10^{12}$. Note also that in each point the parameter $\mathcal{M}$ has been fixed such that we are in agreement with the observed value of the amplitude of the scalar power spectrum $A_s$. Finally, the number of $e$-folds is specified from the reheating constraints of~\eqref{eq:efolds}.
It is worth mentioning that for viable inflation, an upper cutoff on the effective nonminimal coupling appears, $\xi_s \lesssim 4\times 10^{-3}$. This consecutively implies that $ v_s \gtrsim 15\, M_{\rm P}$.

\subsection{Beyond quadratic gravity: Palatini \texorpdfstring{$F(\Ri)$}{F(R)}}
\label{sec:FR}

Let us briefly consider curvature terms beyond the quadratic order. The simplest such models are of the $F(\Ri)$ form, with the action
\begin{equation} \label{eq:SF_Palatini}
S = \int \dd^4 x \sqrt{- g} \left[ \frac{1}{2}F(\Ri) - \frac{1}{2} g^{\mu\nu} \nabla_\mu \varphi \nabla_\nu \varphi - V(\varphi) \right] \, .
\end{equation}
This can be transformed into the Einstein frame following the familiar procedure, first writing
\begin{equation} \label{eq:SF_Palatini_chi}
S_F[g_{\mu\nu},\chi] =  \int \dd^{4}x \sqrt{- g}~ \left[ \frac{1}{2}F'(\chi) (\Ri - \chi) + \frac{1}{2}F(\chi) - \frac{1}{2} g^{\mu\nu} \nabla_\mu \varphi \nabla_\nu \varphi - V(\varphi) \right] \, 
\end{equation}
and then defining $\bar{g}_{\mu\nu} = F'(\chi) g_{\mu\nu}$, so that
\begin{gather} \label{eq:SF_Einstein_Palatini}
S = \int \dd^4 x \sqrt{- \bar{g}} \left[ \frac{1}{2}\bar{R} - \frac{1}{2} g^{\mu\nu} \nabla_\mu \phi \nabla_\nu \phi - U(\phi,\chi) \right] \, , \\
\frac{\dd \varphi}{\dd \phi} = \sqrt{F'(\chi)} \, , \qquad U(\phi, \chi) = \frac{V(\varphi(\phi))}{F'(\chi)^2} - \frac{F(\chi)}{2F'(\chi)^2} + \frac{\chi}{2F'(\chi)} \, .
\end{gather}
The constraint equation for $\chi$ becomes
\begin{equation}
\label{eq:FR_Paltini_constraint}
    4\left[G(\chi) - V(\varphi)\right] - \bar{g}^{\mu\nu}\nabla_\mu \varphi\nabla_\nu \varphi = 0 \, , \quad G(\chi) \equiv \frac{1}{4}\left[2F(\chi)-\chi F'(\chi)\right] \, .
\end{equation}
Unfortunately, this cannot be solved analytically for $\chi$ beyond simple cases, such as the $F(\Ri)=\Ri + \alpha \Ri^2$ discussed above\footnote{The solution is in equation \eqref{eq:constraint_chi}, though note the different conventions for $\chi$ there and in this section.}. However, in the slow-roll limit, we can neglect the derivative term in~\eqref{eq:FR_Paltini_constraint}, obtaining the approximative solution $G(\chi) \approx V(\varphi)$. The slow-roll parameters are easiest to compute in terms of $\chi$ instead of $\varphi$ or $\phi$ using the slow-roll results \cite{Dioguardi:2021fmr}
\begin{equation}
    U(\chi) \approx \frac{\chi}{4F'(\chi)} \, , \qquad \frac{\partial f(\chi)}{\partial \phi} \approx \sqrt{F'(\chi)}\frac{V'(V^{-1}[G(\chi)])}{G'(\chi)} \frac{\partial f(\chi)}{\partial \chi}  \, .
\end{equation}
With these, we can compute the slow-roll CMB predictions~\eqref{eq:power_spectrum_sr}, \eqref{eq:r}. In \cite{Dioguardi:2021fmr}, the procedure was demonstrated for models with $F(\Ri) = \Ri + \alpha\Ri^n$ with multiple values of $n$. Analogously to the standard $n=2$ case, a large $\alpha$ suppresses the tensor-to-scalar ratio $r$.  In \cite{Dioguardi:2022oqu}, models with $F(\Ri) = 2\Lambda - \omega \Ri + \alpha \Ri^2
$ were considered, with a negative coefficient for the linear term ($\omega > 0$), compensated by a negative Jordan frame potential $V$ to make the Einstein frame potential positive-definite.

While such $F(\Ri)$ models may be phenomenologically interesting, comparing them to the $\Ri^2$ case is particularly useful. In~\cite{Dioguardi:2021fmr}, it was shown that models where $F(\Ri)$ grows faster than $\Ri^2$ (so that $G(\chi) < 0$ for large $\chi$) suffer from  multiple problems. In particular, there is a line in the $(\chi,\varphi)$ phase space where the time derivative of $\chi$ diverges, signalling the unhealthiness of such models. This highlights the special nature of the $\Ri^2$ scenario studied above.

For a review of the applications of Palatini $F(\Ri)$ models beyond inflation, see~\cite{Olmo:2011uz}.

\section{Preheating and reheating in Palatini inflation}
\label{sec:preheating}

When inflation ends, the inflaton typically oscillates at the bottom of its potential, decaying into a thermal bath of particles. This process is called \emph{reheating}. Its early stages are typically dominated by \emph{preheating}, the non-perturbative, resonant amplification of field perturbations driven by the time evolution of the inflaton condensate. In this section, we discuss these processes in example Palatini setups.

\subsection{Tachyonic preheating}

Preheating was studied in the context of Palatini models in a series of papers \cite{Rubio:2019ypq, Karam:2020rpa, Karam:2021sno, Tomberg:2021bll, Koivunen:2022mem}, revealing common features. Most of the Einstein frame inflationary potentials discussed in this review are of the plateau form, behaving asymptotically as
\begin{equation} \label{eq:preh_potential}
    U(\phi) \approx U_0\left(1-2n e^{-2|\phi|/\phi_0}\right) \, , \qquad |\phi| \gtrsim \phi_0 \, .
\end{equation}
These models have the standard prediction \eqref{eq:starobinsky_CMB} for $n_s$, and $r\approx 2\phi_0^2/N_\star^2$. At the edge of the plateau, when $\phi \sim \phi_0$, the second derivative of $U$ is highly negative. This causes a tachyonic instability for the field perturbations, which follow the mode equations
\begin{equation} \label{eq:dphi_eom}
    \delta\ddot{\phi}_k + 3H\delta\dot{\phi}_k + \left(\frac{k^2}{a^2} + U''(\phi)\right)\delta\phi_k = 0 \, ,
\end{equation}
where $k$ is the Fourier wave number. When $U''(\phi)<0$, the solutions for $\delta\phi_k$ grow exponentially. Typically, the background inflaton field $\phi$ passes over the tachyonic region quickly, and its energy is diluted by Hubble friction so that it only oscillates near the bottom of the potential, so the tachyonic effect is weak. However, if $\phi_0 \lesssim 0.01$, corresponding to $r\lesssim 10^{-7}$, Hubble friction is inefficient, and the oscillating inflaton returns repeatedly to the tachyonic edge of the plateau~\cite{Tomberg:2021bll}. This leads to an almost instantaneous fragmentation of the condensate, completing preheating in much less than one $e$-folds of expansion. Metric plateau models, such as Starobinsky and Higgs inflation, tend to have a fixed $\phi_0 = \sqrt{6}$, see \eqref{eq:starobinsky_U}, \eqref{eq:higgs_U_metric}, but Palatini models can easily verge into the tachyonic territory, \textit{e.g.} for $\xi \gtrsim 10^4$ in the Higgs model \eqref{eq:higgs_U_Palatini} with $\phi_0 = \xi^{-1/2}$, or for $\alpha \gtrsim 2500 m^{-2} \approx 5\times 10^{13}$ in the $\Ri^2$ quadratic model with \eqref{eq:R2_phi_transformation}, \eqref{eq:R2_quadratic_U}.

Tachyonic preheating in plateau models was first considered for Palatini Higgs inflation in \cite{Rubio:2019ypq}, extended with a sector of supermassive dark matter in \cite{Karam:2020rpa}. In \cite{Karam:2021sno}, the $\alpha\Ri^2$, $V \sim \phi^2$ model from section~\ref{sec:minimal_quadratic} was considered, and it was shown that the higher-order kinetic terms play a negligible role in preheating. Studies \cite{Tomberg:2021bll, Koivunen:2022mem} characterized preheating for the broader class of models \eqref{eq:preh_potential} for small $\phi_0$, providing analytical approximations and charts of the Floquet instability index and commenting on the possibility of producing strong gravitational waves.

Beyond linear analysis, similar models have been studied using lattice techniques \textit{e.g.} in the context of $\alpha$-attractor $T$-models in \cite{Lozanov:2016hid, Lozanov:2017hjm} and explicitly for Palatini Higgs inflation in \cite{Dux:2022kuk}.

\subsection{Reheating temperature}

Let us next consider reheating more generally, beyond the instantaneous tachyonic transition. The number of $e$-folds during the reheating era is given by
\begin{equation}
\label{eq:N_reh}
\Delta N_{\rm reh} \equiv \ln \left[\frac{a_{\rm reh}}{a_{\rm end}}\right] = -\frac{1}{3(1+w)}\ln \left[\frac{\rho_{\rm reh}}{\rho_{\rm end}} \right]\,,
\end{equation}
which vanishes identically for instantaneous reheating, since in this case $a_{\rm reh}= a_{\rm end}$. In terms of $\Delta N_{\rm reh}$, the reheating temperature is given by
\begin{equation}
\label{T_reheat}
T_{\rm reh} = T_{\rm ins} \exp\left(- \frac{3(1+w) \Delta N_{\rm reh}}{4} \right)\,, \quad \text{with} \quad T_{\rm ins} \equiv \left( \frac{30}{\pi^2} \frac{\rho_{\rm end}}{g_\star(T_{\rm reh})}\right)^{1/4}\,.
\end{equation}
$T_{\rm ins}$ is the maximum temperature defined as the instantaneous reheating temperature obtained for $w=1/3$.

Regarding the Palatini inflationary models, the authors of~\cite{Gialamas:2019nly, Das:2020kff, Lykkas:2021vax, Cheong:2021kyc, Lahanas:2022mng, Zhang:2023hfx} studied the inflationary predictions for various values of the equation of state parameter lying in the range $-1/3\le w \le 1$. The resulting reheating temperature, $T_{\rm reh}$, varies between small values close to Big Bang Nucleosynthesis, \textit{i.e.} $T_{\rm reh}\sim \mathrm{MeV}$ up to $T_{\rm reh}\sim 10^{16}\,\mathrm{GeV}$.
Also, the authors of~\cite{Lloyd-Stubbs:2020pvx} studied specific reheating mechanisms, which contain inflaton annihilation to Higgs bosons and reheating via inflaton decay to right-handed neutrinos.  In this review we focus on the computation of the instantaneous reheating temperature in the context of the Palatini inflationary models discussed in subsections~\ref{Example:_Higgs_Inflation} and~\ref{sec:minimal_quadratic}.

In the Palatini Higgs inflationary model of subsection~\ref{Example:_Higgs_Inflation}, the  field value at the end of inflation can be found approximately using the condition $\epsilon_U (\phi_{\rm end})=1$, which gives
$\phi_{\rm end} = \frac{1}{2\sqrt{\xi}}\arcsinh \left(4\sqrt{2}\xi \right)$. Thus, the energy density at the end of inflation is written as\footnote{At the end of inflation, $\epsilon_H = 1$, so $\rho = \frac{1}{2}\dot{\phi}^2 + U = \frac{3}{2}U$, see \eqref{eq:sr_parameters}.}
\begin{equation}
\rho_{\rm end} \simeq \frac{3}{2}U(\phi_{\rm end}) = \frac{3\lambda}{1024\xi^4}\left(16\xi\left(2+8\xi -\sqrt{1+32\xi} \right)-\sqrt{1+32\xi} +1 \right)\,,
\end{equation}
which for $\xi\gg 1$ can be approximated by $\rho_{\rm end}\simeq 3\lambda/(8\xi^2)$. Substituting the latter in equation~\eqref{T_reheat} we obtain that the instantaneous reheating temperature is
\begin{equation}
\label{eq:tins_Higgs}
T_{\rm ins} \simeq 8\times 10^{17} \left(\frac{\lambda}{\xi^2}\right)^{1/4} \mathrm{GeV} \,\simeq \, 10^{13} \,\mathrm{GeV}\,,
\end{equation}
for $\lambda =0.1$ and for $\xi$ being constrained from the amplitude of the scalar power spectrum $A_s$  to the value $\xi \simeq 1.5 \times 10^9$ (in the tachyonic region), for $60$ $e$-folds of inflation. Palatini--Higgs reheating has also been considered in \cite{Fu:2017iqg}.

We will follow the same procedure for the quadratic$+ \mathcal{R}^2$ model described in detail in subsection~\ref{sec:minimal_quadratic}. In this model, the field value at the end of inflation is given by $\phi_{\rm end} =  \frac{\phi_0}{2} \arcsinh (2\sqrt{2}/\phi_0)$, so
\begin{equation}
\rho_{\rm end}  = \frac{3}{16\alpha}\left(1-\frac{\sqrt{1+8/\phi_0^2}-1}{4/\phi_0^2} \right) \Rightarrow\left\{ 
    \begin{array}{lr}
      \rho_{\rm end} \sim \frac{3}{16\alpha} \quad\,  \text{for}\,\,\, \phi_0^2 \gg 1\,,\\[0.2cm]
        \rho_{\rm end} \sim \frac{3m^2}{2} \quad  \text{for}\,\,\, \phi_0^2 \ll 1\,.
    \end{array}
\right.
\end{equation}
Having in mind that the mass parameter is constrained to be $m\simeq 7\times 10^{-6}$ (see equation~\eqref{eq:m_quad}), the large $\phi_0$ limit, $\phi_0^2 \gg 1$, means that $\alpha\ll 10^{10}$, while the other (tachyonic) limit is valid once $\alpha\gg 10^{10}$. In these limits, the instantaneous reheating temperature is
\begin{eqnarray}
\label{Tins:q1}
&&T_{\rm ins} \sim 7\times 10^{17} \alpha^{-1/4} \,\mathrm{GeV} , \quad\,\,\,\,  \text{for}\,\,\, \phi_0^2 \gg 1\\[0.3cm]
&&T_{\rm ins} \sim  10^{18} \sqrt{m} \,\,\mathrm{GeV}, \quad\quad\quad\,\,\, \,\,\, \text{for}\,\,\, \phi_0^2 \ll 1\,,
\label{Tins:q2}
\end{eqnarray}
so using also that $\alpha \gtrsim 10^8$ (see equation~\eqref{eq:al_quad}) we conclude that $T_{\rm ins}\sim 10^{15} - 10^{16} \,\mathrm{GeV} $ in the whole range of the observationally accepted parameter space. 

Reheating in the above model, along with other Palatini-$\Ri^2$ models, has been discussed in more detail in~\cite{Gialamas:2019nly, Lahanas:2022mng}. In~\cite{Lahanas:2022mng}, the higher-order kinetic terms that appear in the EF action~\eqref{eq:action:PalatiniR2:final} have been taken into account, which leads to slightly modified numerical factors in equations~\eqref{Tins:q1} and~\eqref{Tins:q2}.

\section{Palatini \texorpdfstring{$\Ri^2$}{R\^{}2} quintessential inflation}
\label{sec:quintessece}

We next study a Palatini model of \emph{quintessential inflation} from \cite{Dimopoulos:2022tvn, Dimopoulos:2022rdp}, where the inflaton field also produces dark-energy-like late-time acceleration. To study the full evolution of the Universe from early to late times, we add matter fields to the system, denoted below by $\psi$. We choose the action
\begin{equation} \label{eq:S_quintessence}
    S = \int \dd^4 x \sqrt{-g} \left[ \frac{1}{2}F\left(\varphi,\Ri\right) - \frac{1}{2}g^{\mu\nu}\nabla_\mu\varphi\nabla_\nu\varphi - V(\varphi) \right]+S_{\text{m}}[g_{\mu\nu},\psi] \, ,
\end{equation}
with\footnote{Note that the $\alpha$ parameter of \cite{Dimopoulos:2022tvn, Dimopoulos:2022rdp} is, by definition, twice the $\alpha$ used in this paper.}
\begin{equation}
    F(\varphi,\Ri)=\left(1 + \xi \varphi^2 \right)\Ri+\alpha \Ri^2 \,, \qquad \xi (\varphi) = \xi_{*} \left[ 1 + \beta \ln \left( \frac{\varphi^2}{\mu^2} \right) \right] \ .
\end{equation}
The non-minimal coupling runs, starting from $\xi_{*} > 0$ at a reference scale $\mu$ and evolving according to the constant $\beta < 0$.
The inflaton-quintessence field $\varphi$ is governed by an exponential potential
\begin{equation}
    V(\varphi)=M^4 e^{-\kappa\varphi} \, ,
\end{equation}
where $\kappa$ and $M$ are constants.

To understand how this setup leads to quintessential inflation, we transform \eqref{eq:S_quintessence} into the Einstein frame, neglecting the matter fields $\psi$ for now. Following a procedure familiar from the above sections, we get the standard action with a canonical scalar field $\phi$ and the potential
\begin{equation} \label{eq:quintessence_U}
    U(\phi)  = \frac{M^4e^{-\kappa\varphi(\phi)}}{(1 + \xi\varphi(\phi)^2)^2 + 8\alpha M^4e^{-\kappa\varphi(\phi)} } \, , \quad
    \frac{\dd \phi}{\dd \varphi} = \sqrt{\frac{1 + \xi\varphi^2}{(1 + \xi\varphi^2)^2 +8\alpha V(\varphi)}} \, .
\end{equation}
Inflation happens at large negative $\varphi$, where $V$ is high, the $\alpha$ terms dominate in \eqref{eq:quintessence_U}, and $U$ develops a slow-roll plateau. For large positive $\varphi$, the $\alpha$ terms become subdominant, and $U$ exhibits a lower, exponential-like tail, suitable for quintessence. Inflationary behavior is then controlled by $\alpha$, but the $\xi$ term is also needed to tune the plateau to fit the CMB predictions, as explained in detail in \cite{Dimopoulos:2022rdp}. The running $\beta$ allows us to decouple the inflationary $\xi$ from its quintessential value and tune the late-time behavior separately. As we will see below, for reasonable choices of the model parameters, the tail today can have the correct energy density for dark energy, without the extreme fine-tuning of a cosmological constant.

\begin{figure}
    \centering
    \includegraphics[width=0.9\textwidth]{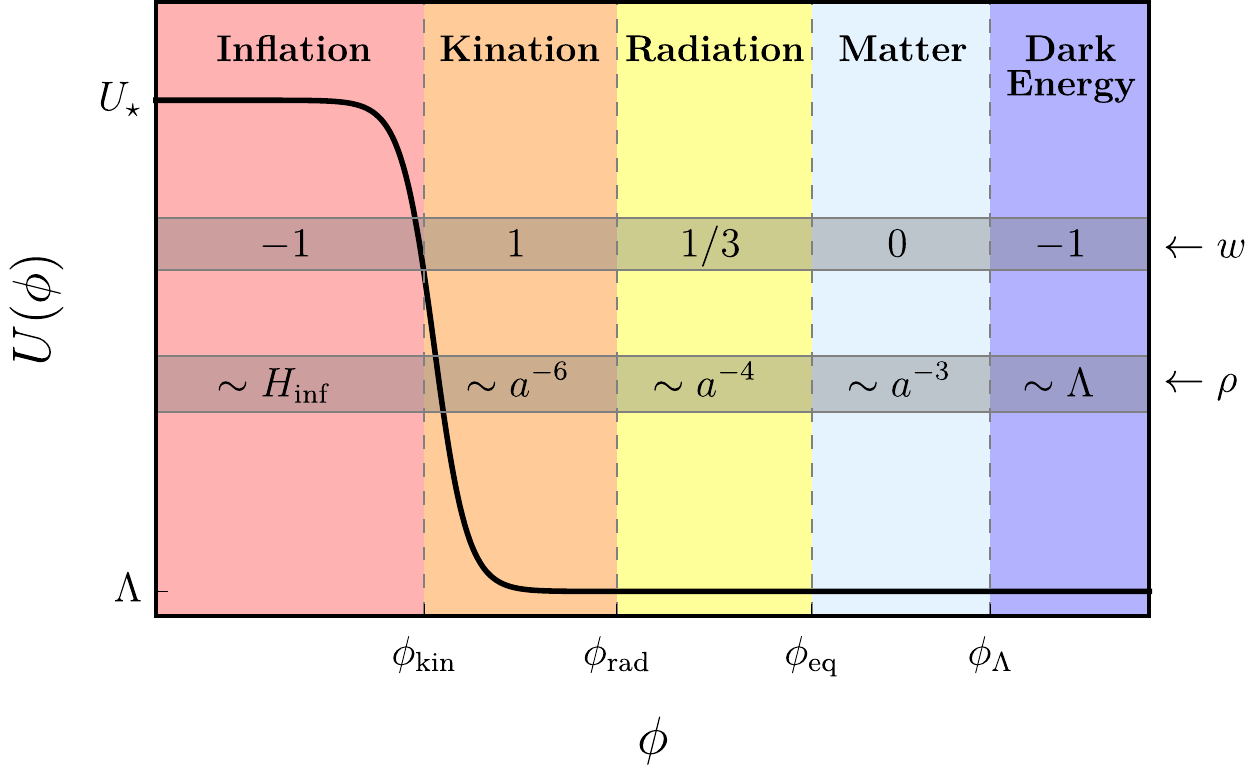}
    \caption{The different phases of quintessential inflation, together with a sketch of the scalar potential. In the notation of section~\ref{sec:quintessece}, $U_\star = 1/(8\alpha)$. $\phi_{\rm kin}$ denotes the value of the field after the end of inflation and during the onset of kination, $\phi_{\rm rad}$ is the field value at the beginning of the radiation era, $\phi_{\rm eq}$ denotes the radiation-matter equality, and $\phi_{\Lambda}$ is the value of the field for which the potential behaves as the cosmological constant.}
    \label{fig:quintessence_phases}
\end{figure}

When the matter degrees of freedom $\psi$ are included, we get the full evolution of the Universe, consisting of a number of phases:
\begin{enumerate}
    \item Inflation for $\phi < 0$, with $\phi$ slowly rolling towards positive values. The energy density $\rho$ is approximately constant, and the barotropic parameter of the Universe  (defined as pressure divided by energy density) is $w=-1$.
    \item Transition to \emph{kination} around $\phi \approx 0$. In kination, energy density is dominated by the kinetic energy of the field, diluting as $\rho\propto a^{-6}$, corresponding to a barotropic parameter $w=1$. Around this time, a small amount of radiation is created, \textit{e.g.} through Ricci reheating \cite{Dimopoulos:2018wfg}.
    \item Since the scalar kinetic energy density dilutes faster than radiation, the latter eventually takes over, leading to the standard radiation domination, where $\rho \propto a^{-4}$, $w=1/3$. The scalar continues rolling. It can follow a tracking solution, mimicking the scaling of the fluid, or freeze to an approximately constant energy density, depending on the potential \cite{Copeland:1997et}. In our setup, the field freezes in a minimum of the Einstein frame potential \eqref{eq:quintessence_U}, induced by the running of $\xi$.
    \item Eventually, the radiation becomes non-relativistic, and matter domination starts, with $\rho \propto a^{-3}$, $w=0$.
    \item The frozen scalar condensate eventually retakes matter, leading to the final dark energy phase with an approximately constant energy density and $w\approx-1$.
\end{enumerate}
The phases are depicted in figure~\ref{fig:quintessence_phases}.
The kination period modifies the standard expansion history of the Universe, pushing the CMB scale to $\sim 70$ $e$-folds before the end of inflation instead of the usual $50$--$60$. Our setup is also complicated by a coupling between the scalar and matter sectors, induced by the Weyl transformation of the metric in $S_{\text{m}}[g_{\mu\nu},\psi]$. This coupling feeds energy from the fluid to the field during the matter-dominated
era, as detailed in \cite{Dimopoulos:2022rdp}.

\subsection{Numerical scan}

In \cite{Dimopoulos:2022rdp}, the coupled scalar-fluid system was solved numerically, and a scan over the parameter space was performed to find points that satisfy all observational constraints. These include constraints for the CMB observables and the dark energy density and its barotropic parameter today. In addition, we demand the initial radiation energy density to be small compared to the field energy density but larger than the minimum bound from gravitational reheating \cite{Ford:1986sy}. We also demand that the temperature of the Universe at the onset of radiation domination is above that of Big Bang nucleosynthesis.

\begin{figure}
    \centering
    \includegraphics[width=0.7\textwidth]{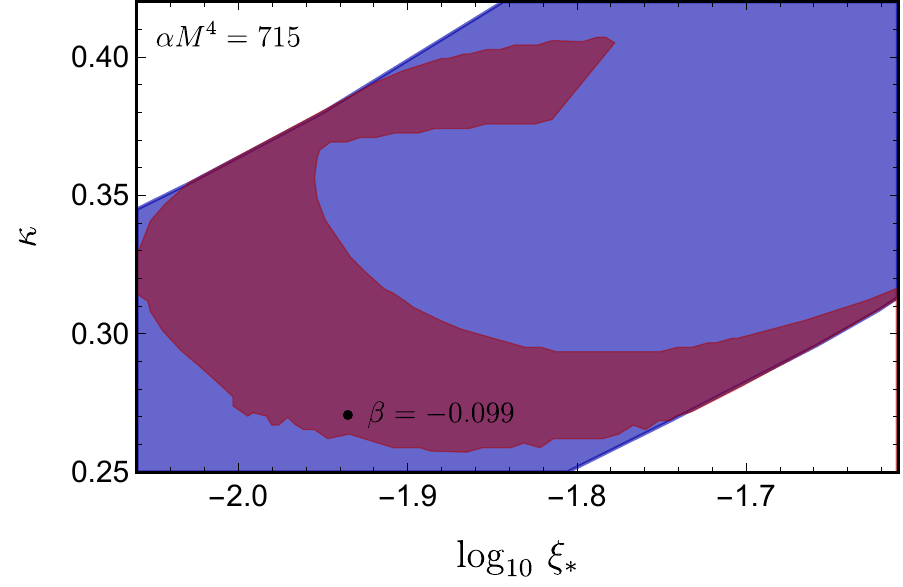}
    \caption{Scan over a slice of the parameter space of the model of section~\ref{sec:quintessece}. The outer blue points satisfy the constraint for $n_s$, while the inner purple points satisfy all CMB constraints. The black point satisfies all observational constraints, including the late-time dark energy requirements. Figure adapted from~\cite{Dimopoulos:2022rdp}.}
    \label{fig:quintessence_scan}
\end{figure}

These constraints are enough to fix almost all of the parameters of the model. Figure~\ref{fig:quintessence_scan} shows a slice of the parameter scan from  \cite{Dimopoulos:2022rdp}. An example scenario satisfying all the constraints is given by
\begin{equation} \label{eq:quintessence_values}
\begin{gathered}
    \alpha = 3.86\times 10^{12} \, , \quad
    \log_{10} \xi_\star = -1.960 \, , \quad
    \beta = -0.100 \, , \\
    \mu = -6 \, , \quad
    M^4 = 1.85\times 10^{-9}\, , \quad
    \kappa = 0.284 \, .
\end{gathered}
\end{equation}

The model presented here is one of the first to produce successful quintessential inflation with modified gravity as the main ingredient. The parameter values \eqref{eq:quintessence_values} are typical, avoiding the fine-tuning of a cosmological constant of $\Lambda \sim 10^{-120}$, instead producing the low energy density dynamically through an exponentially decaying potential. The cosmological evolution deviates in subtle ways from the standard $\Lambda$CDM case, through the aforementioned coupling between $\phi$ and the fluid, and the possibility of a mildly time-dependent dark energy equation of state. Such signatures make quintessence theoretically and phenomenologically interesting.

Quintessential Palatini inflation was also discussed in \cite{Giovannini:2019mgk} (with an $\Ri^2$ term and an emphasis on the post-inflationary stiff phase where $w>1/3$), \cite{Verner:2020gfa} (with a piecewise-defined  Peebles--Vilenkin potential and a non-minimal coupling to gravity), and in \cite{Dimopoulos:2020pas} (a similar Peebles--Vilenkin potential and an $\Ri^2$ term). In \cite{Antoniadis:2022cqh}, a model similar to \cite{Dimopoulos:2022tvn, Dimopoulos:2022rdp} was studied, but with an emphasis on quintessence and no attempt to make the same field produce inflation as well.

\section{Conclusions}

In recent years, there has been a blooming interest in alternative formulations of gravity and their phenomenological ramifications and signatures. In this review, we presented an array of inflationary models studied in the Palatini formalism, where the metric and the connection are assumed to be independent variables, in contrast to the metric formalism where the connection is constrained to be the Levi-Civita one. 

We began by focusing on scalar-tensor theories and saw that the difference between metric and Palatini arises from the transformation of the Ricci scalar and consequently from the form of the non-canonical function multiplying the kinetic term in the Einstein frame. As an example, we studied the model of non-minimally coupled Higgs inflation in both metric and Palatini formalisms. We saw that due to the different field redefinitions, the Einstein frame potential is different and thus the model gives different inflationary predictions in the two formalisms. For $\lambda = 0.1$ and $N_\star = 50$, the expression for the amplitude of the scalar power spectrum gives $\xi \sim 10^4$ in metric and $10^9$ in Palatini, which in turn results in a tensor-to-scalar ratio of $r \sim 10^{-3}$ in metric and $r \sim 10^{-12}$ in Palatini. If future experiments such as PICO, which will be able to probe values of $\delta_r \sim 10^{-4}$, rules out the metric version of the model (and at the same time the Starobinsky model which has the same predictions), then the Palatini version of the model will be strongly favoured. Furthermore, we saw that the literature suggests that tree-level unitarity is violated in the metric but not in the Palatini version of the model, which lends more credibility to the Palatini formalism. 

We then focused on a model containing a non-minimal derivative coupling multiplied by a field-dependent function $m(\varphi)$. We saw that by employing a disformal transformation we can bring the action in the Einstein frame, but a series of higher-order kinetic terms is generated. The field redefinition now also contained the NMDC function. We studied an example with a non-minimally coupled Higgs-like field and a quadratic $m(\varphi) \propto \varphi^2$ function and saw that the value of $r$ can go up compared to the Palatini Higgs inflation model, possibly within reach of future experiments. 

After that, we focused on Palatini quadratic gravity. We saw that the auxiliary field used to eliminate the higher curvature terms is non-dynamical and thus its equation of motion reduces to a constraint. As a consequence, the Einstein frame action is single field and easy to analyse in the context of inflation. However, a new quartic kinetic term is generated and the potential gets modified. The quartic term does not significantly influence the dynamics during or after inflation, but the modified potential which becomes asymptotically flat can greatly impact the value of the tensor-to-scalar ratio (the values of $n_s$ and $A_s$ remain unchanged). In particular, we saw how models which were previously excluded due to their high value of $r$ can be made compatible with observations again if an $\alpha \Ri^2$ term is added and they are studied in the Palatini formalism. Similarly in the $\alpha \Ri^2 + \beta \Ri_{\mu\nu} \Ri^{\mu\nu}$ model endowed with scale invariance and a dynamically-generated Planck scale, we saw how for growing values of the parameter $\tilde{\alpha} = 8 \alpha + 2\beta$ the Coleman-Weinberg potential gets flattened and we can have successful inflation from both sides of the minimum.

We next discussed reheating in models of Palatini inflation. Many of these belong to the class of plateau models with a small tensor-to-scalar ratio, which all exhibit tachyonic preheating, a process where the inflaton quickly fragments into particles, leading to instantaneous reheating. We computed the reheating temperature in the instantaneous case and commented on other possible channels of reheating in Palatini models.

Finally, we applied the Palatini $\Ri^2$ flattening of the inflation potential to a simple model of quintessential inflation. We saw that a single scalar field can drive inflation and then become dynamical dark energy. The numerous experimental constraints heavily reduce the available parameter space of such a model but there are still benchmark points that satisfy all the constraints without the extreme fine-tuning of $\Lambda$CDM. The model is one of the first to produce successful quintessential inflation using modified gravity as the main ingredient and demonstrates the power of the Palatini formalism. 

Despite our understanding of the intricate nuances of the models examined in this paper, several aspects related to Palatini models of inflation and beyond remain inadequately or only partially understood. These include, for instance, multifield scenarios, (p)reheating details in diverse situations, the initial conditions of inflation, quantum corrections to different models, and scenarios with non-zero torsion. Although initial strides have been taken in these directions, it is evident that future investigations hold the potential for revealing more about the association between inflation and various theories of gravity, and hopefully aid us in ultimately identifying the true nature of gravity.

\section*{Acknowledgments}

We thank Antonio Racioppi for useful comments.
This work was supported by the Estonian Research Council grants SJD18, PSG761 and PRG1055 and by the EU through the European Regional Development Fund CoE program TK133 ``The Dark Side of the Universe". TDP acknowledges the support of the Research Centre for Theoretical Physics and Astrophysics of the Institute of Physics at the Silesian University in Opava.

\bibliography{Palatini_Review.bib}

\end{document}